# Atomically Thin Optical Lenses and Gratings

Jiong Yang,[1,†] Zhu Wang,[2,†] Fan Wang,[3] Renjing Xu,[1] Jin Tao,[1] Shuang Zhang,[1] Qinghua Qin,[1] Barry Luther-Davies,[4] Chennupati Jagadish,[3] Zongfu Yu,[2*] and Yuerui Lu[1*]

[1]Research School of Engineering, College of Engineering and Computer Science, the Australian National University, Canberra, ACT, 0200, Australia

[2]Department of Electrical and Computer Engineering, University of Wisconsin, Madison, Wisconsin 53706, USA

[3]Department of Electronic Materials Engineering, Research School of Physics and Engineering, the Australian National University, Canberra, ACT, 0200, Australia

[4]CUDOS, Laser Physics Centre, Research School of Physics and Engineering, the Australian National University, Canberra, ACT 0200, Australia

[†] These authors contributed equally to this work

[*] To whom correspondence should be addressed: Zongfu Yu (zyu54@wisc.edu) and Yuerui Lu (yuerui.lu@anu.edu.au)

**Two-dimensional (2D) materials have emerged as promising candidates for miniaturized optoelectronic devices[1-9], due to their strong inelastic interactions with light[10,11]. On the other hand, a miniaturized optical system also requires strong elastic light-matter interactions to control the flow of light[12]. Here, we report giant optical path length (OPL) from a single-layer molybdenum disulfide ($MoS_2$), which is around one order of magnitude larger than that from a single-layer graphene. Using such giant OPL to engineer the phase front of optical beams, we demonstrated, to the best of our knowledge, the world's thinnest optical lens consisting of a few layers of $MoS_2$ less than 6.3 nm thick. Moreover, we show that $MoS_2$ has much better dielectric response than good conductor (like gold) and other dielectric materials (like Si, $SiO_2$ or graphene). By taking advantage of the giant elastic scattering efficiency in ultra-thin high-index 2D materials, we**

**demonstrated high-efficiency gratings based on a single- or few-layers of MoS$_2$. The capability of manipulating the flow of light in 2D materials opens an exciting avenue towards unprecedented miniaturization of optical components and the integration of advanced optical functionalities.**

Interactions between light and matter can be divided into two categories: inelastic and elastic[12]. An inelastic interaction involves energy transfer between photons and electrons or phonons. In contrast, elastic interactions do not involve energy transfer, and are responsible for controlling the propagation of light. Optical components, such as resonant cavities, waveguides, lenses, gratings, and, more recently, optical meta-materials[13] and photonic crystals[14], all rely on strong elastic interactions between light and matter to achieve sophisticated control of the flow of light. Strong elastic interactions rely on significant changes of the amplitude and phase of the light accumulated over a long optical path and, hence, for very thin materials, such as a 2D graphene sheet, the interaction is generally very small[15]. Considerable effort has been devoted to this issue, but success has been only achieved in the mid- to far-infrared where the plasmonic resonance in graphene can enhance the elastic optical response[16-18]. It remains a great challenge to manipulate the flow of light using atomically thin 2D materials in the important visible and near-infrared spectral regions where 2D materials have most interesting optoelectronic properties. Rather surprisingly, as we will show later, the strength of the elastic interaction in a thin 2D material increases dramatically with increasing refractive index because of the unique geometry associated with an ultra-thin film. Such favorable scaling makes high-index transition-metal dichalcogenide (TMD) 2D semiconductors[4,19-22], such as MoS$_2$, particularly attractive for strong elastic light-matter interactions.

Refractive optical components rely on the optical path length (OPL) to modify the phase front of an optical beam. The OPL is directly related to the geometrical length of light path. As a result, it is normally expected that the OPL of an ultra-thin 2D material would be too small to have a significant impact on the phase front because of their ultra-thin thicknesses. Here we have been able to observe a giant OPL of 38 nm from a single-layer MoS$_2$, which is more than 50 times larger than its physical thickness of 0.67 nm and around one order of magnitude larger than the measured OPL of a single-layer of graphene that was found to be only 4.4 nm (Figure 1).

In our experiments, single- or few-layer MoS$_2$ flakes were transferred onto an silicon wafer with 275 nm of surface thermal oxide by mechanical exfoliation[15,18]. The flakes were firstly identified by their optical contrast in an optical microscope. Regions with different colors corresponded to MoS$_2$ flakes with different thicknesses (Figure 1a). Due to their high refractive index, these atomically thin MoS$_2$ layers have significant and layer-dependent OPL values and this enables the layers to be easily identified by phase-shifting interferometry (PSI) (Figure 1b&c). PSI is capable of measuring the vertical OPL to an accuracy of around 0.1 nm, by analyzing the digitized interference pattern obtained during a well-controlled phase shift (Supplementary information, Figure S1&S2). The measured OPL value of the MoS$_2$ flake on a SiO$_2$ substrate at 535 nm was determined by $OPL_{MoS_2} = -\frac{\lambda}{2\pi}(\phi_{MoS_2} - \phi_{SiO_2})$, where $\lambda$ is the wavelength of the light source, $\phi_{MoS_2}$ and $\phi_{SiO_2}$ are the PSI measured phase shifts of the light reflected from the MoS$_2$ flake and the SiO$_2$ substrate (Figure 1d inset), respectively. We characterized multiple samples and obtained statistical values of the OPL for single- and few-layer MoS$_2$ samples as shown in Figure 1d. For each number of layers of MoS$_2$, at least five different samples were characterized in these statistical measurements. The layer number could be quickly determined by the measured layer-dependent OPL values and the deduced layer number was confirmed by corresponding atomic force microscopy (AFM) images (Figure 1e&f); Raman microscopy (Figure S3a); and photoluminescence (PL) measurements (Figure

S3b) on the same samples. For comparison, we performed the same characterizations on mechanically exfoliated graphene samples (Figure 1d and Figure S4, S5&S6). The measured average and standard deviation error of the OPL values from 1L, 2L, 3L, and 4L $MoS_2$ samples were (38.0 ± 2.8) nm, (85.4 ± 2.2) nm, (124.0 ± 6.6) nm, and (162.6 ± 9.0) nm, respectively, while those from 1L, 2L, 3L, and 4L graphene samples were (4.4 ± 0.8) nm, (8.2 ± 2.0) nm, (13.0 ± 3.2) nm, and (17.2 ± 3.6) nm, respectively, indicating that $MoS_2$ has an OPL per layer approximately an order of magnitude larger than graphene.

This giant OPL is created by relatively strong multiple reflections at the air-$MoS_2$ and $MoS_2$-$SiO_2$ interfaces. We consider a simple interface between air and $SiO_2$, each occupying half infinite space. A layer of 2D material with a real refractive index $n$ is placed in between the two media. The high impedance mismatch at these interfaces leads to large reflection coefficients, which cause the strong multiple reflections of light in the 2D material (Figure 2a). The amplitude of the reflected light is the summation of the multiple reflections off the interfaces of the thin high index layer $R_i$, where $i$ indicates the number of round trips in the 2D material. As the index increases so does the reflectivity of the interfaces, which increases the effective number of transits of the light through the high index layer and thus the OPL of the reflected light (Figure 2a). We verify this intuition with numerical calculation as shown by the dashed line in Figure 2c. The magnitude of OPL difference comparing with and without the 2D material on $SiO_2$ (Supplementary information, Figure S7) increases rapidly with increasing $n$. The OPL is low for low-index 2D materials, where the small reflection coefficients cause $R_i$ to be small. This situation in illustrated schematically in Figure 2b. Additionally, in the experiment, we used a silicon substrate with a layer of 275 nm thermal $SiO_2$ on its surface, which forms a weak Fabry-Perot resonance. As a result of this weak resonant enhancement, the OPL is further enhanced by a factor of around 1.5 as shown by the solid line in Figure 2c. Figure 2c also shows the OPL for a few other materials. The OPL of high-index 2D materials, such as MoS$_2$, is remarkably larger than that of SiO$_2$, graphene, Au or Si. The wavelength used for these calculations was 535 nm. The refractive indices used for MoS$_2$[22], silicon, graphene[23], SiO$_2$ and Au were 5.3+1.3$i$, 4.15+0.0439$i$, 2.6+1.3$i$, 1.46, and 0.467 + 2.4$i$, respectively. In addition, it should be noted that the giant OPL is not a narrow band effect. The calculated OPL for 1L MoS$_2$ is above 20 nm at the wavelength ranging from 450 nm to 560 nm (Supplementary information, Figure S8). The spectral position for highest OPL can be adjusted by changing the thickness of the SiO$_2$.

Even more remarkably, the OPL of single-, bi-, triple- and quadri-layer MoS$_2$ scales almost linearly with the number of layers, offering the exciting opportunity of controlling the OPL using a number of layers of MoS$_2$. When the layer thickness increases by 1 nm, the OPL increases by over 50 nm. Such a rapid change of OPL with thickness allows us to control the phase front of an optical beam very effectively using only an atomically thin structure. The theoretical and numerical predictions (Figure 2d) were well supported by the experimental data as shown in (Figure 1d).

Next, we demonstrate phase-front engineering by fabricating the world's thinnest lens based on a few atomic layers of MoS$_2$ (Figure 3). We started with a flake of uniform 9L MoS$_2$ (6.28 nm in thickness, Figure S9) and then used a focused ion beam (FIB) to mill a pre-designed bowl-shape structure (20 μm in diameter) into the flake (Figure 3a&b). The gradual change of MoS$_2$ thickness, from the center to the edge, led to a continuous and curved OPL profile for an incident beam, and this served as an atomically thin (reflective) concave micro-lens (Figure 3c). Based on the measured OPL profile, the focal length $f$ of this MoS$_2$ micro-lens was calculated to be -248 μm (Supplementary information, Figure S10). In order to realize the precise design for this MoS$_2$ micro-lens, we used the statistical calibration curve between the OPL values of MoS$_2$ flakes and their layer numbers (Figure 3d). All the OPL values were measured by PSI and the layer numbers were confirmed by AFM. The OPL of $MoS_2$ increased almost linearly with increasing the layer number when the layer number was less than five.

We used a far-field scanning optical microscopy (SOM) to characterize the fabricated $MoS_2$ micro-lens (Supplementary information, Figure S11). The SOM system used a green laser (at 532 nm) that was focused onto the focal plane of an Olympus 10X (NA = 0.25, depth of focus 18 μm) objective lens. The setup offered the best collection efficiency for light emitted from a small volume located around the focal plane. The micro-lens was moved along the $z$-axis in steps of 10 μm by a piezo-electrically driven stage. The camera recorded a series of the intensity distributions (Figure S12) with the $MoS_2$ micro-lens positioned at different $z$ values. A three-dimensional dataset was generated by data processing and a cross sectional profile was obtained along the $x$- and $z$-axes to illustrate the average distribution of the light intensity in these directions (Figure 3e). When the $MoS_2$ micro-lens was placed at a distance $2|f|$ above the focal plane, the focused incident light would be exactly reimaged which is equivalent to the light coming from a point source (Figure S12d). Therefore, the camera recorded a well-focused light spot. The focal length $f$ of the $MoS_2$ micro-lens was measured to be -240 μm ($2f$ = -480 μm), which matched very well with the simulated value (-248 μm) using the measured OPL profile of the micro-lens. For comparison, we also ran the same characterization by using a planar substrate without the $MoS_2$ micro-lens, and obtained the intensity distribution shown in Figure 3f and Figure S13. The lensing effect is clearly demonstrated by comparing the difference between Figure 3e and 3f. In addition, the measured focal length of the $MoS_2$ micro-lens shows weak polarization dependence (Figure S14), due to the low anisotropic dielectric response of $MoS_2$. This makes $MoS_2$ suitable for ultra-thin optical elements.

The efficiency of light scattering is another critical parameter for advanced light manipulation. Devices that employ photonic band gaps[24], Anderson localization[25], and light trapping such as with thin-film solar cells[26] all rely heavily on strong light scattering. Unfortunately, in typical 2D materials, such as graphene, the scattering efficiency is very small, making it impossible to rely on collective scattering of nanostructured graphene to achieve functionalities such as gratings. Here, we show that single- and few-layer structured MoS$_2$ film have extraordinarily high scattering efficiency, enabled by the combination of high index in a thin structure. The scattering efficiency is determined by the strength of the electric field in the material. Normally, the electric field inside a bulk material, particularly a high-index material is much weaker than that of incident light because of the impedance mismatch. The boundary condition of Maxwell's equations requires the tangential component of the electric field to be continuous across any interface. Because the layer is thin, this condition indicates that the electrical field inside a 2D material is almost as strong as the tangential component of the incident field. As a result, there is a strong polarization $P = \epsilon_0(n^2 - 1)E_0$, where $E_0$ is the electric field of s-polarized incident light, *n* is the index of the material and $\epsilon_0$ is the electric permittivity of free space. The scattering power is proportional to the $P^2$ and, therefore, scales roughly as $n^4$. This scaling rule greatly favors high-index materials and is again uniquely available in ultra-thin materials. In contrast, for nanoparticles, the scattering power is proportional to $\left(\frac{n^2-1}{n^2+2}\right)^2$, which does not increase appreciably with the refractive index[27].

Here we use finite element method to explicitly calculate the scattering efficiency of 2D ribbons by solving Maxwell's equations. Figure 4a shows the calculated scattering cross section of an infinitely long ribbon (30 nm wide and 0.67 nm thick) in air for s-polarized light incident from the normal direction. The scattering cross section has units of nanometers because the length of ribbon is considered infinite. The scattering cross section increases by orders of magnitude when the index increases by just a few times (Figure 4a). For example, the scattering cross section of a single-layer MoS$_2$ ribbon is around 670 times, 54 times and 18 times of those in 0.67 nm $SiO_2$, a single-layer graphene and 0.67 nm of gold, respectively. Metal is generally considered as one of the strongest scattering materials and it is important to note that $MoS_2$ even displays much stronger light scattering than gold. Moreover, the angular response of the scattering cross section is also isotropic (Figure S15). Such favorable scaling for high-index materials is uniquely available in ultra-thin materials. The giant scattering efficiency in high-index 2D materials makes it possible to achieve sophisticated light manipulation based on collective scattering by nanostructured patterns. Next, we experimentally demonstrate efficient optical gratings made from only a few layers of atoms. Because of the giant scattering efficiency, the efficiency of $MoS_2$ gratings is orders of magnitude greater than those made from conventional materials, such as $SiO_2$ and gold, and other low-index 2D materials.

We used FIB to mill grating patterns on 1L, 2L, 6L and 8L $MoS_2$ flakes (Figure 4 and Figure S16, S17&S18). Grating parameters used in experiments, such as the periodicity and filling ratio, were based on optimal configuration predicted by simulations (Supplementary information, Table S1). The gratings were characterized using an s-polarized green laser (at a wavelength of 532 nm). The laser beam has a diameter of around 200 μm, which was large enough to fully cover the grating. First-order and second-order diffraction beams were observed and the measured diffraction angles agreed with the predictions of the diffraction equation $d(sin\theta_d + sin\theta_i) = m\lambda$, where $\theta_d$ and $\theta_i$ are the diffraction angle and incident angle respectively; *d* is the period of the grating elements; and *m* is an integer characterizing the diffraction order. The power of the first-order diffraction beam was measured and the grating efficiency $\eta$ was determined by $\eta = (P_d/P_i) * (S_b/S_g)$, where $P_d$ and $P_i$ were the measured powers of the diffracted and incident beams, respectively; $S_b$ and $S_g$ were the measured areas of the incident beam and the $MoS_2$ grating, respectively. The measured grating efficiency is a function of the incident angle, which agrees well with our simulation (Figure

4g). The maximum grating efficiencies for the 1L, 2L, 6L and 8L $MoS_2$ gratings were measured to be 0.3%, 0.8%, 4.4% and 10.1%, respectively, which also agree well with the simulations (Figure 4h, Table S1). For comparison, we also fabricated a grating from a graphene sheet deposited by large-area chemical vapor deposition (Supplementary information, Figure S19a&b). The intensity of diffracted beam from the graphene grating was lower than the noise level of our light detection system, and thus had a maximum efficiency no greater than 0.02%. From our simulations, the maximum grating efficiency of mono-layer graphene would be only 0.0078%, which is around 47 times lower than that of a single-layer $MoS_2$ grating. As another comparison, a $SiO_2$ grating with 2 nm thickness was also fabricated (Figure S19c&d). Again no diffracted beam could be observed from the $SiO_2$ grating due to the low grating efficiency in accordance with our numerical predictions (Figure 4h, Table S1).

The efficiency of the $MoS_2$ grating can be further improved by using a metallic mirror to replace the Si substrate. Based on simulations of optimized designs, the $1^{st}$ order grating efficiency of an 8L $MoS_2$ grating can be up to 23.7% (Table S2, Figure S20&21). In addition, an asymmetrical profile as used in high-efficiency gratings is expected to further improve the efficiency.

In conclusion, we have shown that high-index 2D materials have extraordinary elastic interactions with light, enabled uniquely by the ultra-thin nature of 2D materials. As a result, wavefront shaping[28,29] and efficient light scattering can be accomplished with atomically thin 2D materials, enabling a new class of optical components entirely based on high-index 2D materials. Moreover, compared to conventional diffractive optical components, the spatial resolution of phase-front shaping is much smaller than the wavelength, and is only limited by the nano-fabrication resolution, making it possible to eliminate undesired diffractive orders[29]. 2D materials also offer many unique advantages. First, considering the strong tunability of 2D materials, advanced beam steering can be envisioned[29]. Secondly, we also observed similar giant OPL in other TMD family $YX_2$ (Y=Mo, W; X=S, Se, Te) semiconductors, such as $WS_2$ and $WSe_2$ (Figure S22). The availability of different functional materials offer rich opportunities for the combination of optical and electronic properties, such as stacked atomically thin heterostructures for 2D optoelectronics. Thirdly, high-quality 2D TMD semiconductors can be deposited directly onto (or transferred to) various substrates with large size by chemical vapor deposition at low cost[30] potentially enabling low-cost flexible optical components. Lastly, 2D optical components represents a significant advantage in manufacturing compared to conventional 3D optical components, because different functionalities can all be achieved in a 2D platform sharing the same fabrication processes and this will greatly facilitate the large-scale manufacturing and integration. In summary, our work here opens an exciting opportunity to use high-index 2D materials to control the flow of light.

**Methods**

**Device Fabrication and Characterization.** Single- and few-layer TMD semiconductors and graphene for the PSI measurements were deposited onto a $SiO_2$/Si substrate (275 nm thermal $SiO_2$) by mechanical exfoliation using 3M scotch tape. All Raman and PL measurements were conducted with a Horiba Jobin Yvon T64000 micro-Raman/PL system, with a 532 nm green laser for excitation. All the OPL characterizations were obtained using a phase-shifting interferometer (Vecco NT9100). The atomically thin micro-lens and gratings were fabricated in an FEI FIB system (Gallium ion source) using pre-calibrated dosage, optimized beam voltage (30 kV) and beam current (9.7 pA). The gratings and micro-lens were characterized using a green laser with a wavelength of 532 nm.

**Numerical Simulation.** Rigorous Coupled-Wave Analysis (RCWA) was used to calculate the phase delay and grating efficiency. The method numerically solves Maxwell's equations in multiple layers of structured materials by expanding the field in the Fourier-space. The finite element method was used to calculate the optical scattering cross section of the nano ribbons.

**Author Contributions**

Y.R.L. and Z.F.Y. designed the project; J. Y., R.J.X., S. Z. carried out sample mechanical exfoliation and microscope imaging; J. Y. carried out the OPL, Raman and PL measurements, AFM imaging, grating and micro-lens fabrication, grating efficiency measurement and micro-lens characterization, with partial assistance from Y.R.L; Z. W. and Z.F.Y. conducted the simulations and grating/micro-lens designs; F.W. and C. J. built the optical characterization set up for gratings and micro-lens. B. L-D. set up the PSI measurement system and provided technical support for the OPL characterization. J. T. and Q.H.Q. undertook data processing for the micro-lens images. All authors contributed to the manuscript.

**Acknowledgements**

We would like to acknowledge support from the ACT node of the Australian National Fabrication Facility (ANFF) and, particularly, the technical support on the FIB provided by Dr Li (Lily) Li. We also thank Professor Jin-cheng Zheng, from Xiamen University (China), for helpful discussions, and Professor Vincent Craig and A/Professor Lan Fu, from the Australian National University, for support on AFM and optical imaging. We acknowledge the financial support from an ANU PhD scholarship; the Office of Naval Research (USA) under grant number N00014-14-1-0300; the Australian Research Council (grant number DE140100805); and the ANU Major Equipment Committee.

**Competing financial interests**

The authors declare that they have no competing financial interests.

**FIGURE CAPTIONS**

**Figure 1 | Giant optical path lengths (OPLs) from single- and few-layer MoS₂. a,** Optical microscope image of a mechanically exfoliated $MoS_2$ sample on a $SiO_2$/Si substrate (275 nm thermal $SiO_2$). Different contrasts correspond to $MoS_2$ flakes of different thicknesses. The areas labeled as "1L", "2L", "3L"and "4L" are single-, bi-, triple- and quadruple-layer $MoS_2$, respectively. **b,** Phase shifting interferometry (PSI) image of the region inside the box indicated by the dashed line in **(a)**. **c,** PSI measured OPL values versus position for 1L, 2L, 3L and 4L $MoS_2$ along the dashed line in **(b)**. **d,** Statistical data of the OPL values from PSI for 1L, 2L, 3L and 4L $MoS_2$ and graphene samples. For each layer number of $MoS_2$ and graphene, at least five different samples were characterized for the statistical measurements. Inset is the schematic plot showing the PSI measured phase shifts of the reflected light from the $MoS_2$ flake ($\phi_{MoS_2}$) and the $SiO_2$ substrate ($\phi_{SiO_2}$). **e,** Atomic force microscopy (AFM) image of 1L and 2L $MoS_2$ from the box enclosed by the dashed line 1 in **(b)**. **f,** AFM image of 3L and 4L $MoS_2$ from box enclosed by the dashed line 2 in **(b)**.

**Figure 2 | High refractive index enabled giant OPL in ultra-thin film. a-b,** Schematic plots of multiple reflections at the interfaces of ultra-thin 2D materials. High refractive index leads to a large reflection coefficient. Light is reflected many times inside the material and leads to a highly enhanced light path, indicated as (**a**). For low refractive index material, the light path is much less enhanced because of the small reflection coefficient, indicated as (**b**). **c,** Simulated OPL values for light reflected from 2D material (0.67 nm in thickness) with different indices on a $SiO_2$ (275 nm)/Si substrate (solid line) and $SiO_2$ substrate with infinite thickness (dashed line). The calculated OPLs of 0.67 nm Au, 0.67 nm $SiO_2$, 1L (0.34 nm) graphene and 1L (0.67 nm) $MoS_2$ are represented by markers. **d,** Simulated OPL values for 1L, 2L, 3L and 4L $MoS_2$ and graphene on $SiO_2$ (275 nm)/Si substrate, respectively. This deviates slightly from the linear

relation obtained in experiments as shown in Figure 1d because the refractive index values for different layers are expected to be slightly different whilst constant index values were used for all simulations. The wavelength used in the simulations was 535 nm.

**Figure 3 | Atomically thin micro-lens fabricated from a few-layers of MoS$_2$. a,** PSI image of an atomically thin micro-lens fabricated on a 9L MoS$_2$ flake. **b,** Schematic plot of the micro-lens structure. The bowl-shape structure of the micro-lens was fabricated by focused ion beam (FIB) milling with atomic resolution in the vertical direction and sub-20 nm resolution in lateral direction. **c,** Measured OPL values versus position for the direction indicated by the dashed line in (a). **d,** Measured statistical data of the OPL values for MoS$_2$ flakes with different layers ranging from 1L to 11L. For each layer number of MoS$_2$, at least five different samples were characterized in the statistical measurements. All the layer numbers were confirmed by AFM. **e,** Intensity distribution pattern of the MoS$_2$ micro-lens measured by scanning optical microscopy (SOM). **f,** Intensity distribution pattern of the planar reference SiO$_2$/Si substrate measured by the same SOM setup.

**Figure 4 | Atomically thin high-efficiency gratings made from a single- and a few-layers of MoS$_2$. a,** Simulated scattering cross section (SCS) versus refractive index for a layer of 0.67 nm thick material. The SCS values of 0.67 nm Au, 0.67 nm SiO$_2$, 1L (0.34 nm) graphene and 1L (0.67 nm) MoS$_2$ are represented by markers. The dashed line from equation $y = 0.000167 * x^4$ is added as a reference. **b,** Schematic of the setup of the grating and for measurement of its diffraction efficiency. **c-d,** Optical microscope images of 1L and 2L, and 8L MoS$_2$ gratings. **e-f,** AFM images of 1L and 2L, and 8L MoS$_2$ gratings. Note: based on the measured grating height, the 1L, 2L and 8L MoS$_2$ were fully etched through and the SiO$_2$ substrates underneath were over etched by around 1.5 nm. From our control SiO$_2$ grating experiments and from simulations, the grating contribution from this over etched SiO$_2$ is negligible. **g,** Simulated and measured efficiency of an 8L MoS$_2$ grating versus incident angle

of the light beam. **h,** The comparison of the simulated and measured maximum grating efficiencies for different materials. The dash line represents the noise level of our light detection system, with the minimum detectable grating efficiency being 0.02%.

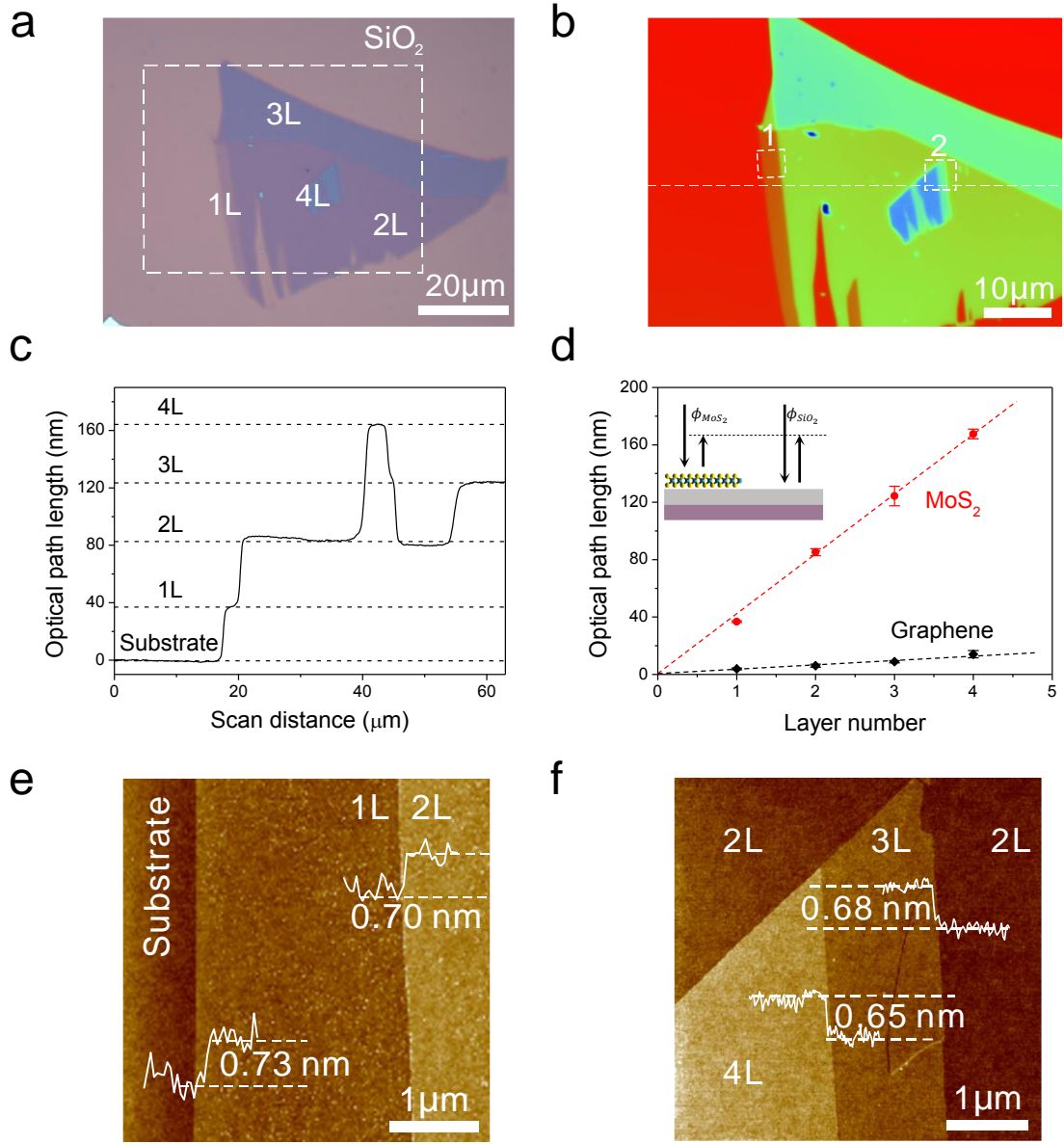

Figure 1

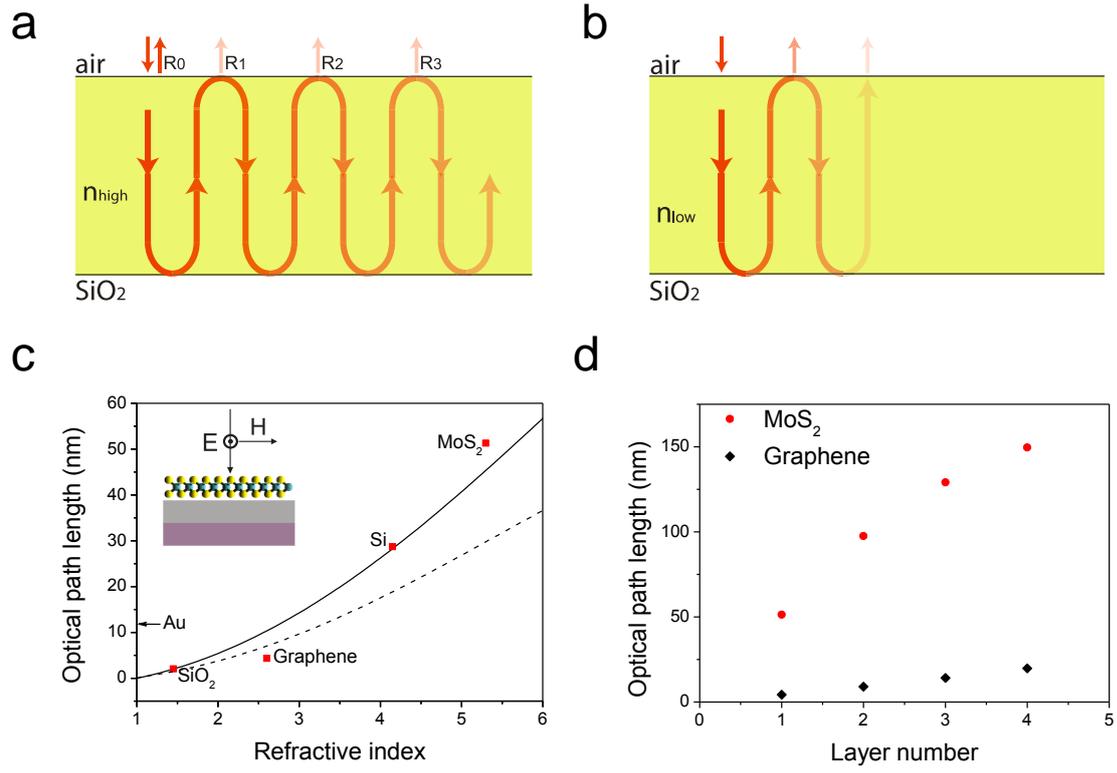

Figure 2

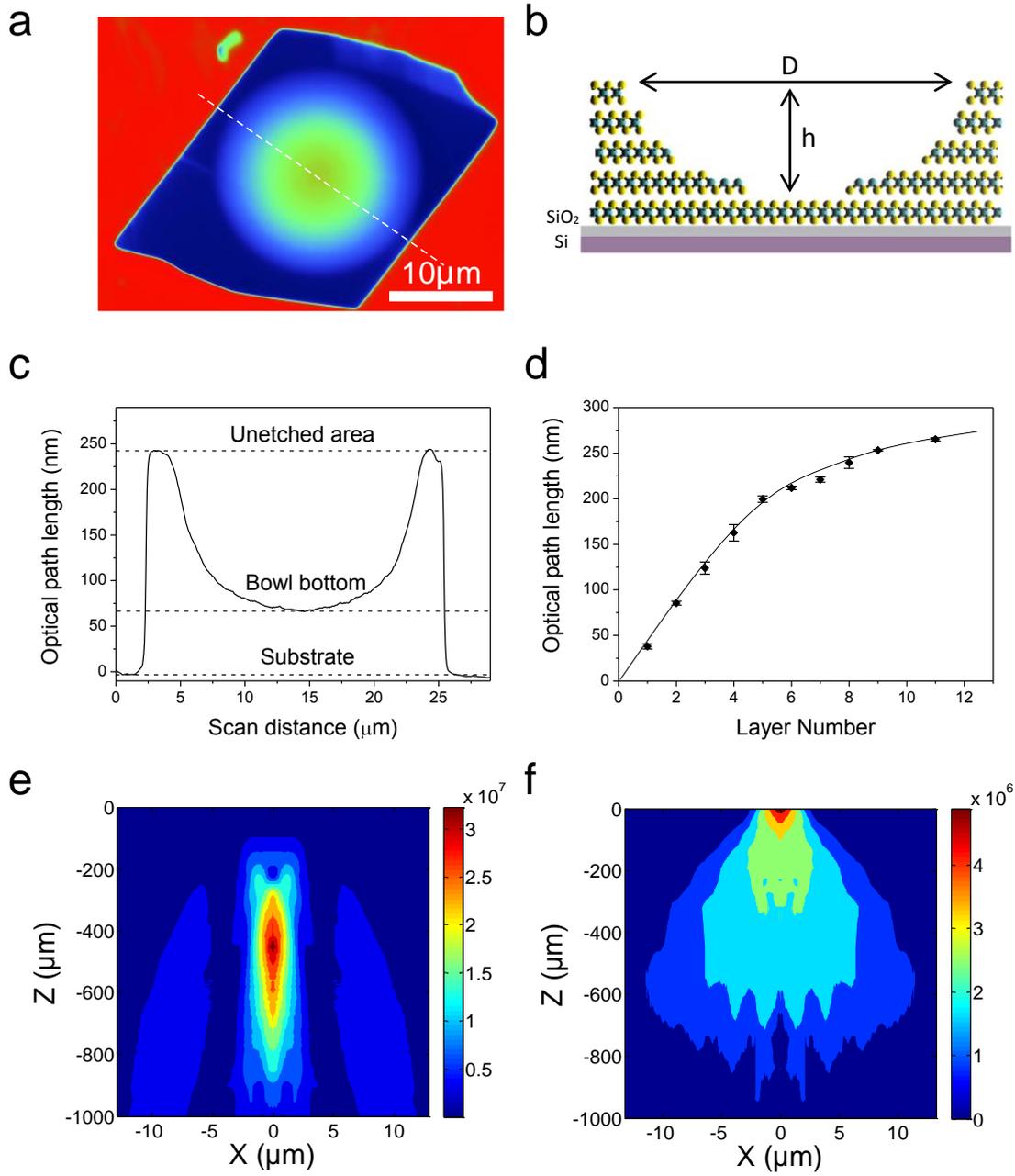

Figure 3

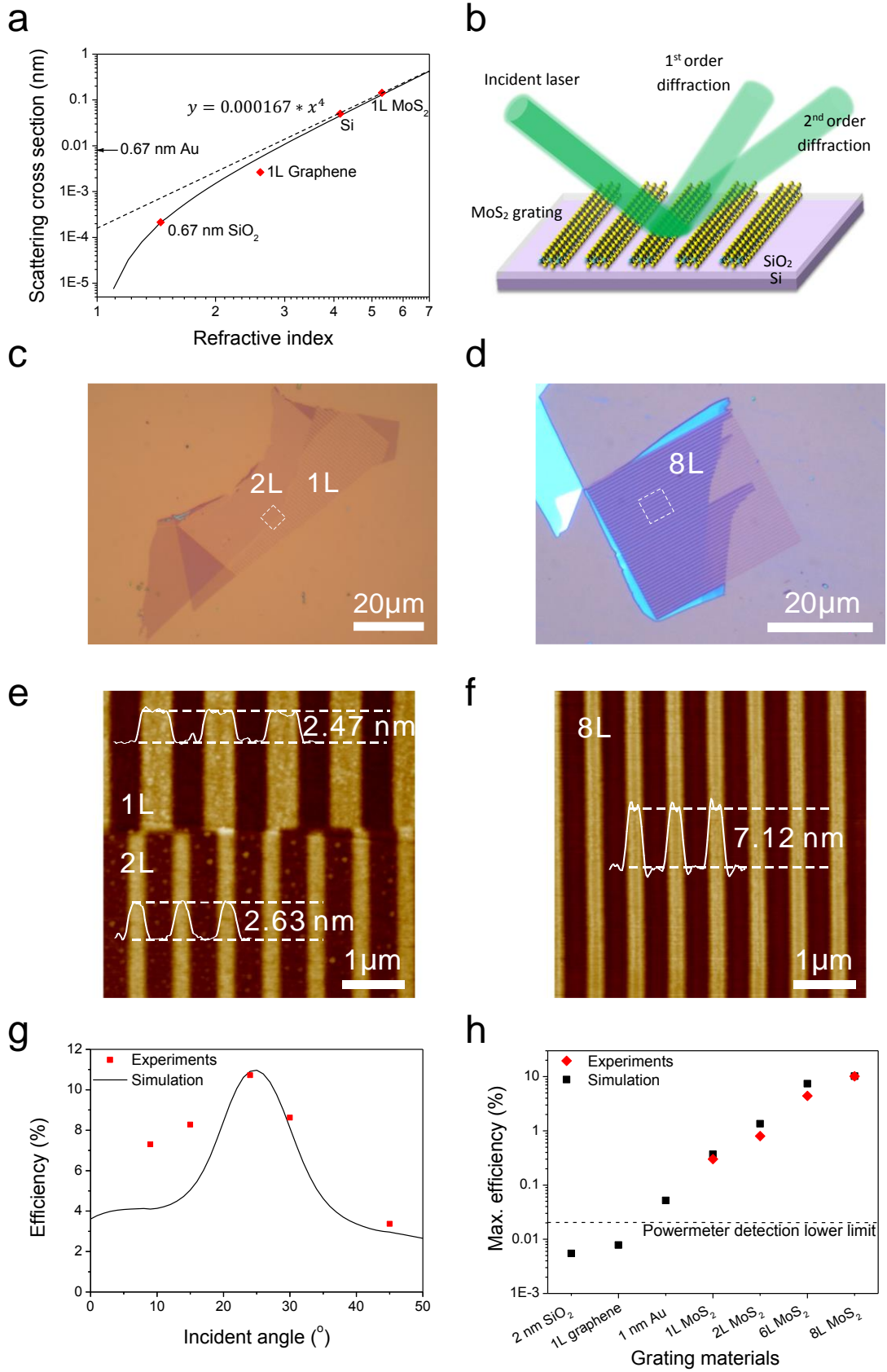

Figure 4

**Supplementary Information for**

# Atomically Thin Optical Lenses and Gratings

Jiong Yang,[1†] Zhu Wang,[2†] Fan Wang,[3] Renjing Xu,[1] Jin Tao,[1] Shuang Zhang,[1] Qinghua Qin,[1] Barry Luther-Davies,[4] Chennupati Jagadish,[3] Zongfu Yu,[2*] and Yuerui Lu[1*]

[1]Research School of Engineering, College of Engineering and Computer Science, the Australian National University, Canberra, ACT, 0200, Australia

[2]Department of Electrical and Computer Engineering, University of Wisconsin, Madison, Wisconsin 53706, USA

[3]Department of Electronic Materials Engineering, Research School of Physics and Engineering, the Australian National University, Canberra, ACT, 0200, Australia

[4] CUDOS, Laser Physics Centre, Research School of Physics and Engineering, the Australian National University, Canberra, ACT 0200, Australia

[†] These authors contributed equally to this work

[*] To whom correspondence should be addressed: Zongfu Yu (zyu54@wisc.edu) and Yuerui Lu (yuerui.lu@anu.edu.au)

**1. Sample preparation for single- and few-layers of graphene, $MoS_2$, $WS_2$ and $WSe_2$**

The bulk graphite crystal was purchased from SPI Supplies®; the bulk $MoS_2$, $WS_2$ and $WSe_2$ crystals were purchased from HQ Graphene. The thin-layers of graphene, $MoS_2$, $WS_2$ and $WSe_2$ were mechanically exfoliated onto a $Si/SiO_2$ substrate (the $SiO_2$ layer was 275 nm thick) using 3M scotch tape, similar to the technique described by other researchers[1-3]. After exfoliation, the thin-layers of graphene, $MoS_2$, $WS_2$ or $WSe_2$ were located with a Leica optical microscope. The physical thickness and layer number of the thin graphene, $MoS_2$, $WS_2$ and $WSe_2$ layers were all confirmed with a Bruker III atomic force microscope (AFM) in tapping mode.

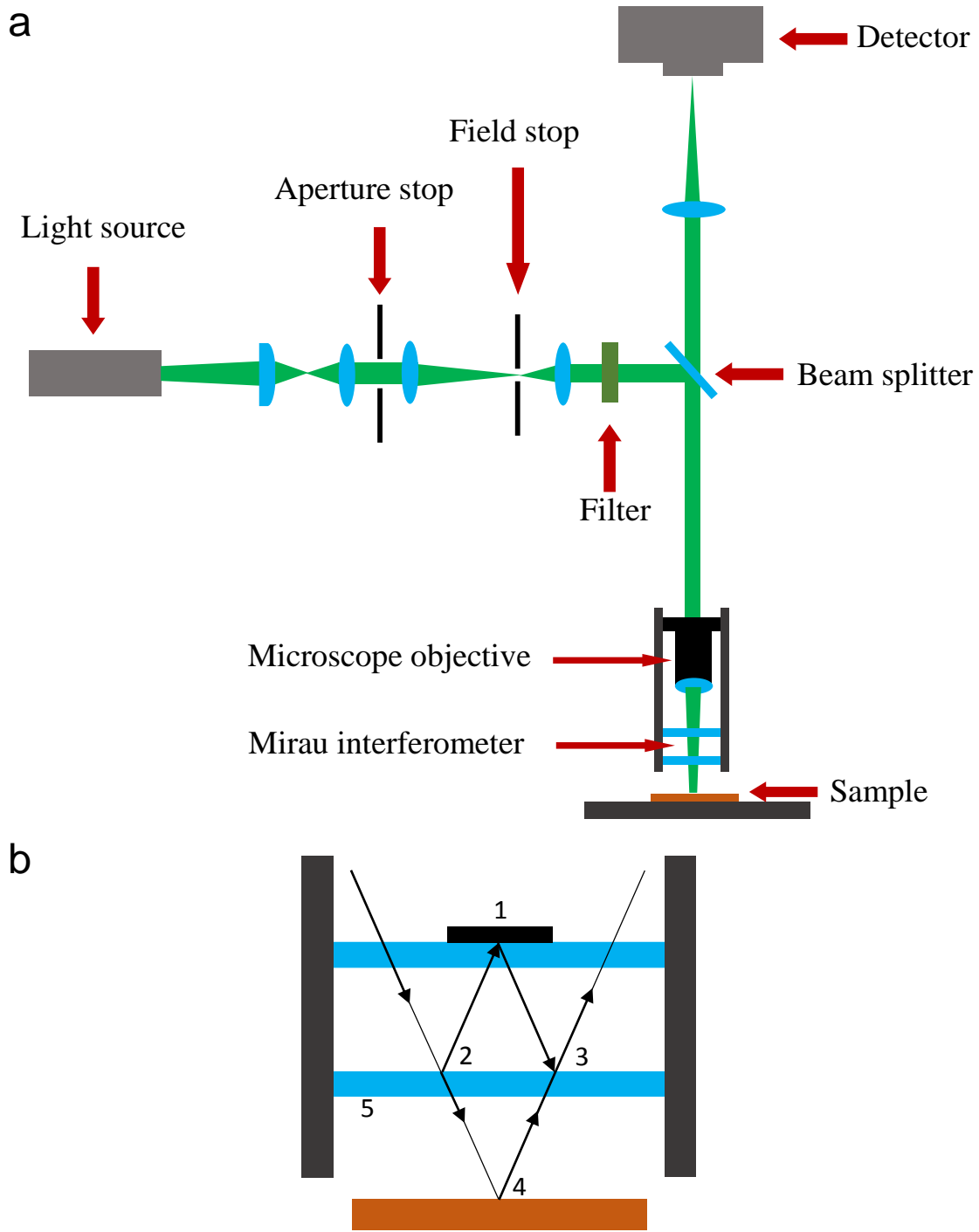

**Figure S1 | Schematic plot of the phase shifting interferometry (PSI) system. a,** Schematic plot of the PSI system. **b,** Zoomed view of the Mirau interferometer. 1. Reference mirror; 2. First reflection of the reference beam; 3. Third reflection of the reference beam; 4. Reflection of the test/objective beam; 5. Semi-transparent mirror. 2-1-3 represents the reference beam and 2-4-3 represents the test/objective beam.

## 2. Phase-shifting interferometry (PSI) working principle

PSI was used to investigate the surface topography based on analyzing the digitized interference data obtained during a well-controlled phase shift introduced by the Mirau interferometer[4]. The PSI system (Vecco NT9100) used in our experiments operates with a green LED source centered near 535 nm by a 10 nm band-pass filter[5]. The schematic of the PSI system is shown in Figure S1a.

The working principle of the PSI system is as follows[6]. For simplicity, wave front phase will be used for analysis. The expressions for the reference and test wave-fronts in the phase shifting interferometer are:

$$w_r(x,y) = a_r(x,y)e^{i\phi_r(x,y)} \tag{S1}$$

$$w_t(x,y,t) = a_t(x,y)e^{i[\phi_t(x,y)+\delta(t)]} \tag{S2}$$

where $a_r(x,y)$ and $a_t(x,y)$ are the wavefront amplitudes, $\phi_r(x,y)$ and $\phi_t(x,y)$ are the corresponding wavefront phases, and $\delta(t)$ is a time-dependent phase shift introduced by the Mirau interferometer. $\delta(t)$ is the relative phase shift between the reference beam and the test beam.

The interference pattern of these two beams is:

$$w_i(x,y,t) = a_r(x,y)e^{i\phi_r(x,y)} + a_t(x,y)e^{i[\phi_t(x,y)+\delta(t)]} \tag{S3}$$

The interference intensity pattern detected by the detector is:

$$I_i(x,y,t) = w_i^*(x,y,t) * w_i(x,y,t) = I'(x,y) + I''(x,y)\cos[\phi(x,y) + \delta(t)] \tag{S4}$$

where $I'(x,y) = a_r^2(x,y) + a_t^2(x,y)$ is the averaged intensity, $I''(x,y) = 2a_r(x,y) * a_t(x,y)$ is known as intensity modulation and $\phi(x,y)$ is the wavefront phase shift $\phi_r(x,y) - \phi_t(x,y)$.

From the above equation, a sinusoidally-varying intensity of the interferogram at a given measurement point as a function of $\delta(t)$ is shown below:

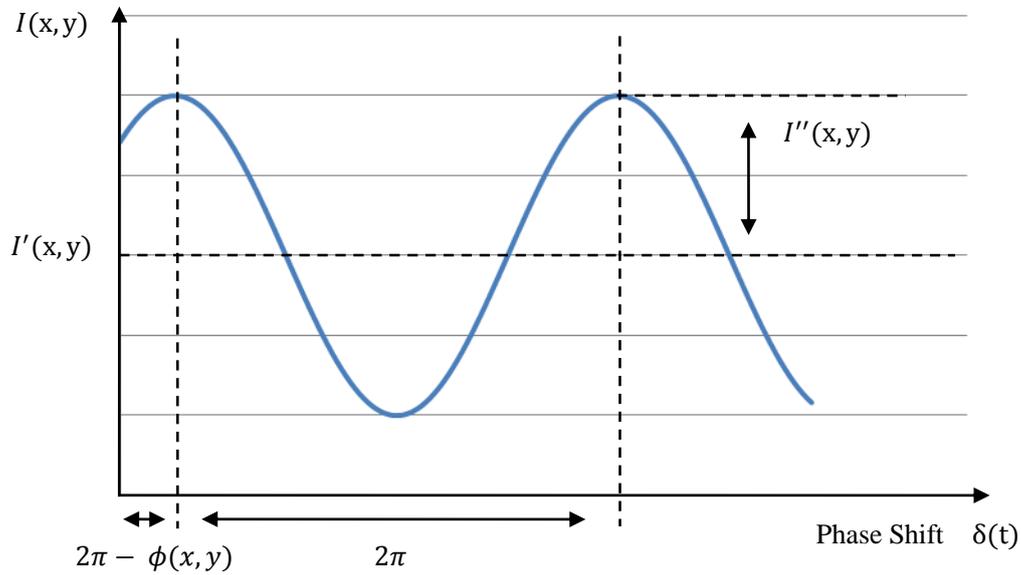

**Figure S2 | Variation of intensity with the reference phase at a point in an interferogram.** $I'(x,y)$ is the averaged intensity, $I''(x,y)$ is half of the peak-to-valley intensity modulation and $\phi(x,y)$ is the temporal phase shift of this sinusoidal variation.

$\delta(t)$ is introduced by the Mirau interferometer, which is shown in Figure S1. When the Mirau interferometer gradually moves toward the sample platform, the optical path length (OPL) of the test beam decreases while the OPL of the reference beam remains invariant.

The computational method of PSI is a four-step algorithm, which needs to acquire four separately recorded and digitalized interferograms of the measurement region. For each separate and sequential recorded interferograms, the phase shift difference is:

$$\delta(t_i) = 0, \frac{\pi}{2}, \pi, \frac{3\pi}{2}; \quad i = 1,2,3,4 \tag{S5}$$

Substituting these four values into the equation S4, leads to the following four equations describing the four measured intensity patterns of the interferogram:

$$I_1(x,y) = I'(x,y) + I''(x,y)\cos[\phi(x,y)] \tag{S6}$$

$$I_2(x,y) = I'(x,y) + I''(x,y)\cos[\phi(x,y) + \frac{\pi}{2}] \tag{S7}$$

$$I_3(x,y) = I'(x,y) + I''(x,y)\cos[\phi(x,y) + \pi] \tag{S8}$$

$$I_4(x,y) = I'(x,y) + I''(x,y)\cos[\phi(x,y) + \frac{3\pi}{2}] \tag{S9}$$

After the trigonometric identity, this yields:

$$I_1(x,y) = I'(x,y) + I''(x,y)\cos[\phi(x,y)] \tag{S10}$$

$$I_2(x,y) = I'(x,y) - I''(x,y)\sin[\phi(x,y)] \tag{S11}$$

$$I_3(x,y) = I'(x,y) - I''(x,y)\cos[\phi(x,y)] \tag{S12}$$

$$I_4(x,y) = I'(x,y) + I''(x,y)\sin[\phi(x,y)] \tag{S13}$$

The unknown variables $I'(x,y)$, $I''(x,y)$ and $\phi(x,y)$ can be solved by only using three of the four equations; but for computational convenience, four equations are used here. Subtracting equation S11 from equation S13, we have:

$$I_4(x,y) - I_2(x,y) = 2I''(x,y)\sin[\phi(x,y)] \tag{S14}$$

And subtract equation S12 from equation S10, we get:

$$I_1(x,y) - I_3(x,y) = 2I''(x,y)\cos[\phi(x,y)] \tag{S15}$$

Taking the ratio of equation S14 and equation S15, the intensity modulation $I''(x,y)$ will be eliminated as following:

$$\frac{I_4(x,y) - I_2(x,y)}{I_1(x,y) - I_3(x,y)} = \tan[\phi(x,y)] \tag{S16}$$

Rearranging equation S16 to get the wave-front phase shift term $\phi(x,y)$:

$$\phi(x,y) = \tan^{-1}\frac{I_4(x,y) - I_2(x,y)}{I_1(x,y) - I_3(x,y)} \tag{S17}$$

This equation is performed at each measurement point to acquire a map of the measured wave-front. Also, in PSI, the phase shift is transferred to the surface height or the optical path difference (OPD):

$$h(x,y) = \frac{\lambda\phi(x,y)}{4\pi} \tag{S18}$$

$$OPD(x,y) = \frac{\lambda \phi(x,y)}{2\pi} \qquad (S19)$$

Here, the OPL of the MoS₂ flake is calculated as:

$$OPL_{MoS_2} = -(OPD_{MoS_2} - OPD_{SiO_2}) = -\frac{\lambda}{2\pi}(\phi_{MoS_2} - \phi_{SiO_2}) \qquad (S20)$$

where $\lambda$ is the wavelength of the light source, $\phi_{MoS_2}$ and $\phi_{SiO_2}$ are the measured phase shifts of the reflected light from the MoS₂ flake and the SiO₂ substrate, respectively. In our experiments, $\phi_{SiO_2}$ was typically set to be zero, as shown in Figure 1c.

### 3. Raman and photoluminescence (PL) measurements

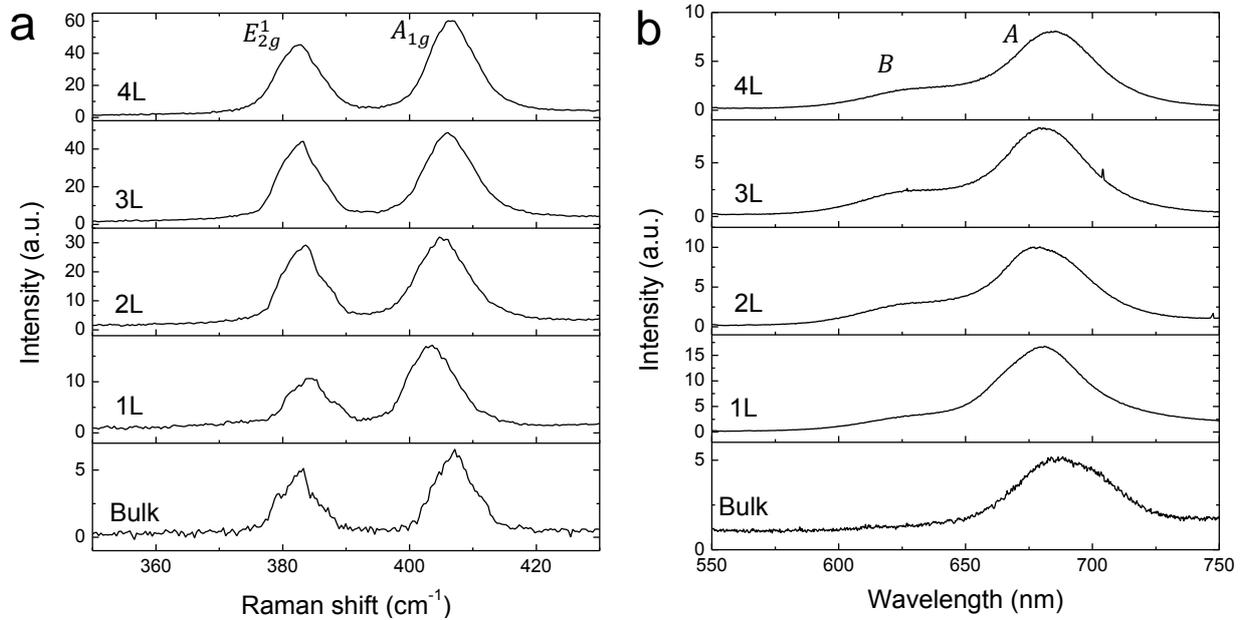

**Figure S3 | Raman and PL spectra of the MoS₂ flakes. a-b,** Raman spectra (**a**) and PL spectra (**b**) of 1L, 2L, 3L, 4L and bulk MoS₂.

All Raman and PL measurements were recorded with a Horiba Jobin Yvon T64000 micro-Raman/PL system, with 532 nm green laser excitation. Figure S3a shows the Raman spectra of our MoS₂ sample containing 1L, 2L 3L and 4L, and the Raman spectrum of a very thick piece (indicated as "bulk"). Two Raman phonon modes $E_{2g}^1$ (in-plane vibrations) and $A_{1g}$ (out-of-plane vibrations) were observed in all layers. With the increase of the layer number, the $E_{2g}^1$ peak shows a red shift, while $A_{1g}$ peak shows a blue shift, consistent with the results of other researchers[7,8]. Figure S3b shows the PL spectra of our MoS$_2$ sample containing 1L, 2L, 3L and 4L, and the PL spectrum of a very thick piece (indicated as "bulk"). Peak A (located at around 625 nm for 1L MoS$_2$) and B (located at around 680 nm for 1L MoS$_2$) were observed and indicated on the spectra, similar to previously reported data[3,9]. These two peaks are associated with excitonic transitions at the *K* point of the Brillouin zone and the energy difference between these two peaks can be attributed to the degeneracy breaking of the valence band due to the spin-orbit coupling[9]. These Raman and PL spectra measurement further confirm the identification of the layer numbers.

4. Images and characterization of graphene

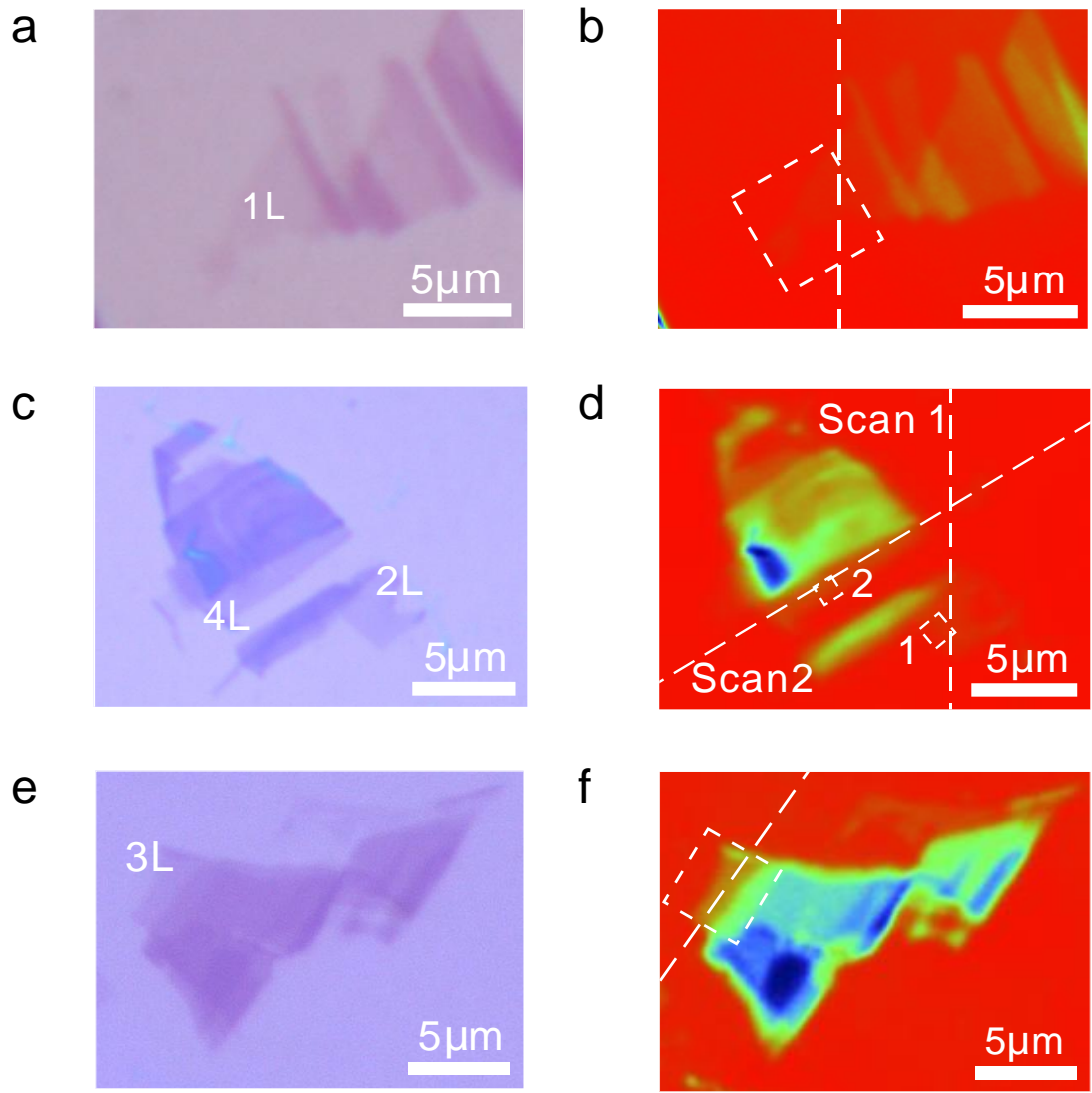

**Figure S4 | Optical microscope and PSI images of mechanically exfoliated graphene. a, c** and **e** display the optical microscope images of 1L, 2L&4L, and 3L graphene, respectively. **b, d** and **f** display the PSI images of the same area in **(a), (c)** and **(e)**, respectively.

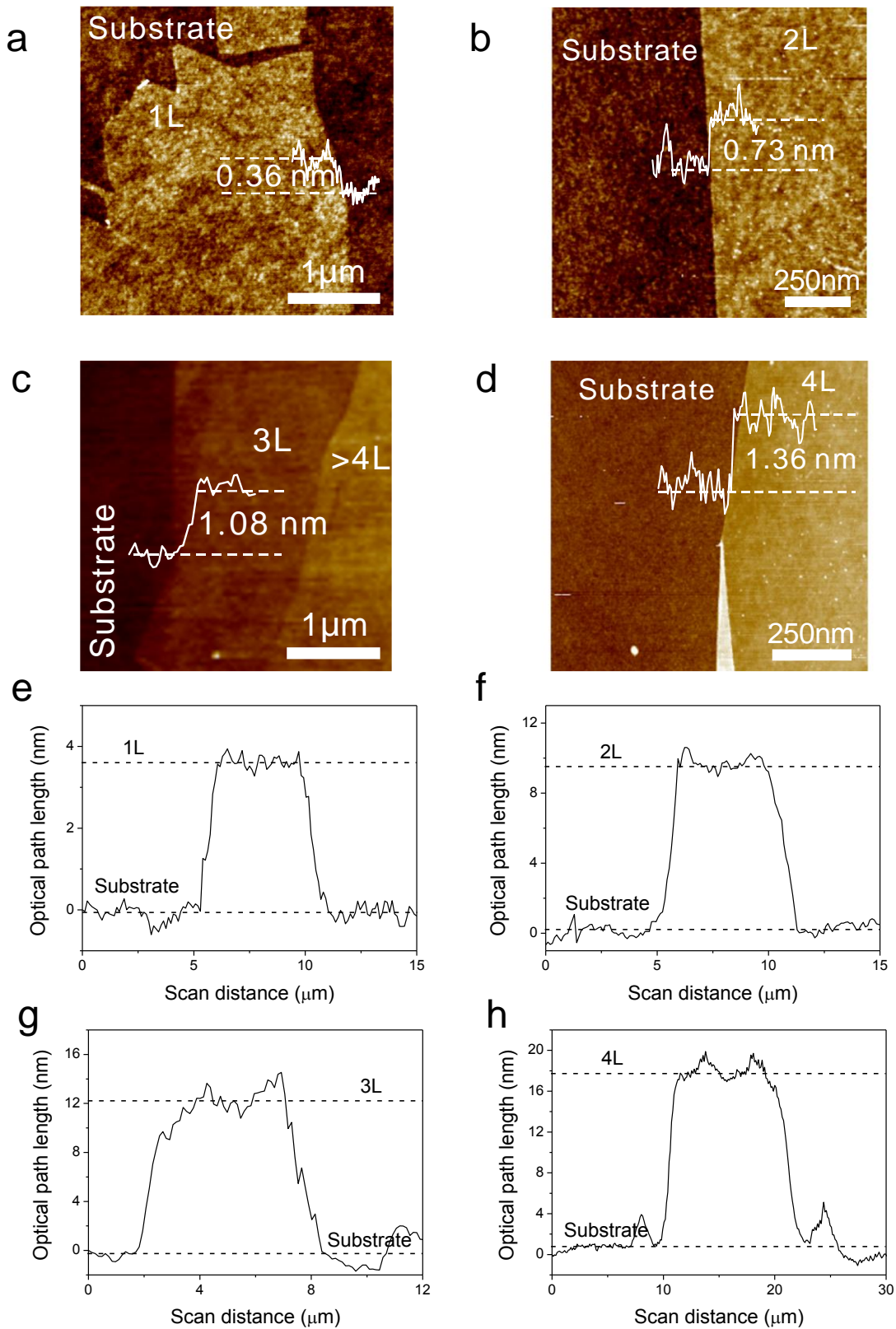

**Figure S5 | AFM images of graphene flakes and their measured OPLs by PSI. a-d,** AFM images of 1L to 4L graphene indicated in the dash line boxes in Figure S4b, S4d and S4f. **e-h,** OPLs of 1L to 4L graphene along the dash lines in Figure S4b, S4d and S4f.

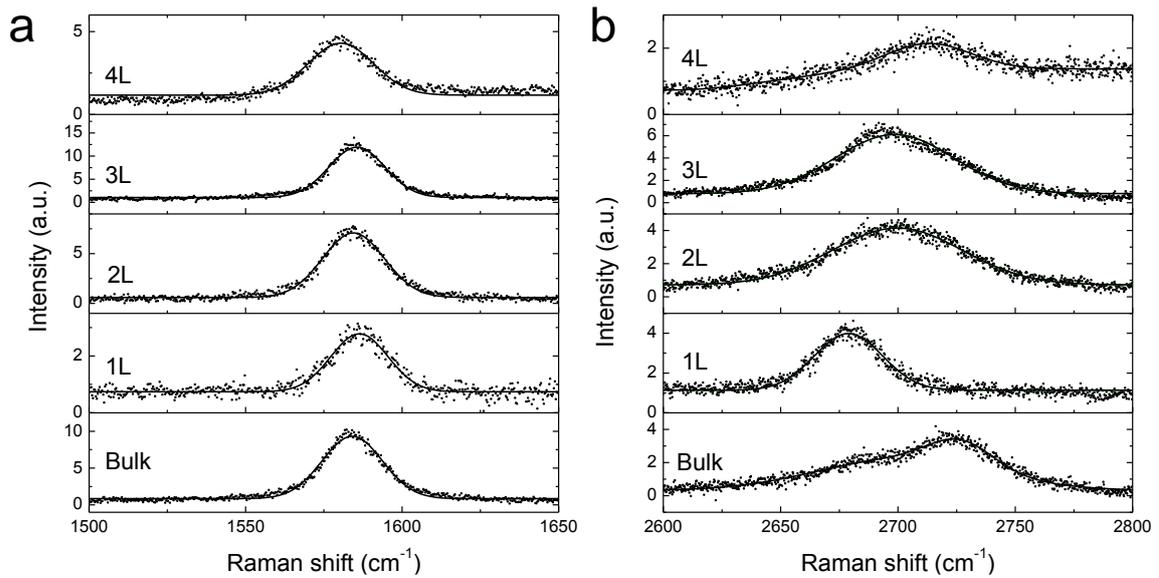

**Figure S6 | Raman spectra of graphene. a-b,** Raman spectra showing the *G* band (**a**) and *G'* (**b**) band from 1L to 4L graphene and bulk graphite samples.

## 5. Calculations for the optical path length (OPL) of atomically thin 2D materials

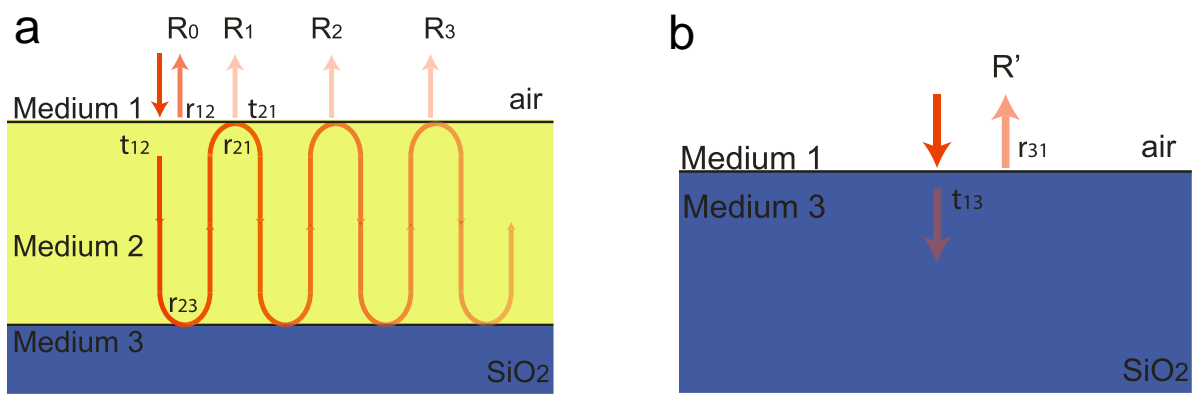

**Figure S7 | a**, Reflection of a three-layer structure. Medium 1 is air, Medium 2 is the 2D material and Medium 3 is an infinite $SiO_2$ substrate. **b**, The reference configuration. Light is incident from air into infinite $SiO_2$ substrate.

The incident light comes from the air resonates inside the 2D material. The total reflection is determined by the interference of all reflected beams $R_i$. To calculate the amplitude of the total reflection, we use $r_{ij}$ ($i,j=1,2,3$) to represent the reflection coefficients when light goes from medium $i$ to medium $j$.

$$r_{ij} = \frac{n_i - n_j}{n_i + n_j} \tag{S21}$$

We use $t_{ij}$ ($i,j$ =1,2,3) to represent the transmission from medium $i$ to medium $j$

$$t_{ij} = \frac{2n_i}{n_i + n_j} \tag{S22}$$

where $n_i$, $n_j$ ($i,j$ =1,2,3) is the refractive index of medium $i,j$. Assuming that the thickness of the 2D material is $d$ and wave vector of incident light in air is $k_0$, we can calculate the reflection of each order,

$$\begin{aligned} R_0 &= r_{12} \\ R_1 &= t_{12} r_{23} t_{21} e^{i2k_0 nd} \\ R_2 &= t_{12} r_{23} r_{21} r_{23} t_{21} (e^{i2k_0 nd})^2 \\ R_3 &= t_{12} r_{23} r_{21} r_{23} r_{21} r_{23} t_{21} (e^{i2k_0 nd})^3 \end{aligned} \tag{S23}$$

where $2k_0 nd$ is the round trip propagation phase and $n$ is the refractive index of the 2D material.

Then the total reflected amplitude is the summation of all reflections, which is

$$\begin{aligned} R &= R_0 + R_1 + R_2 + \\ &= r_{12} + t_{12} r_{23} t_{21} e^{i2k_0 nd} \left[ 1 + r_{21} r_{23} e^{i2k_0 nd} + \left( r_{21} r_{23} e^{i2k_0 nd} \right)^2 + \cdots \right] \\ &= r_{12} + \frac{t_{12} r_{23} t_{21} e^{i2k_0 nd}}{1 - r_{21} r_{23} e^{i2k_0 nd}} \\ &= \frac{1-n}{1+n} + \frac{4n}{(1+n)^2} \frac{(n-1.46)}{(n+1.46)} e^{i2k_0 nd} \frac{1}{1 - \frac{(n-1)(n-1.46)}{(n+1)(n+1.46)} e^{i2k_0 nd}} \end{aligned} \tag{S24}$$

Here we used refractive indices of air and $SiO_2$ as 1 and 1.46, respectively.

The OPL was calculated by comparing the phase difference of the reflected light with and without the 2D material. Figure S7b shows the reference setup. Light is incident directly from air into infinite $SiO_2$ substrate. In this case the reflected amplitude is

$$R' = \frac{n_1 - n_3}{n_1 + n_3} \tag{S25}$$

So we get:

$$OPL = -\frac{\bigl(phase(R)-phase(R')\bigr)}{2\pi}\lambda \qquad (S26)$$

where $\lambda$ is the wavelength of light. The magnitude of OPL is plotted in Figure 2.

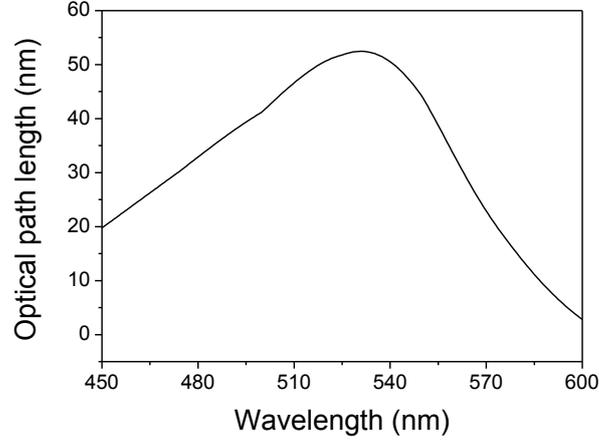

**Figure S8** | Calculated OPL as a function of wavelength for 1L MoS$_2$ placed on a substrate consisting of 275 nm SiO$_2$ on a Si wafer. The wavelength dependent dielectric constant is obtained from reference[10].

We also calculated the OPL as a function of wavelengths for 1L MoS$_2$ as shown in Figure S8. The OPL is above 20 nm for wavelengths ranging from 450 nm to 560 nm with a bandwidth of around 100 nm. We can adjust the thickness of SiO$_2$ layer to shift this operational bandwidth to any desired wavelength range.

## 6. Atomically thin micro-lens fabrication and characterization

### 6.1 Micro-lens fabrication

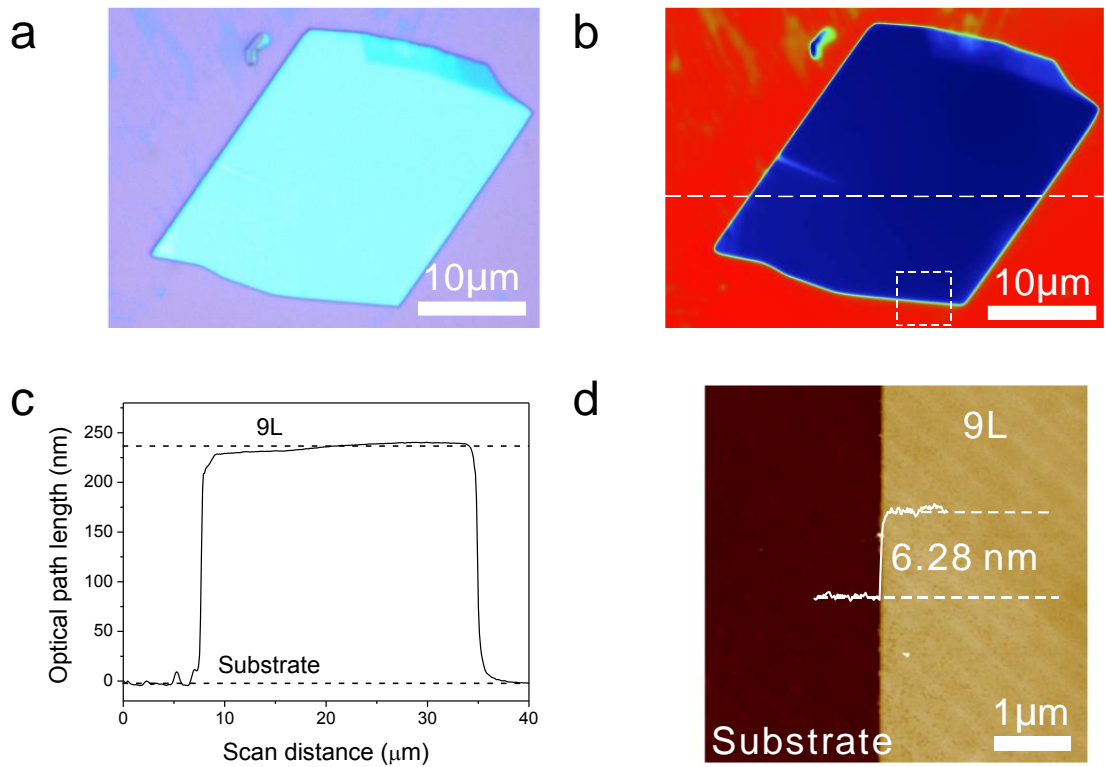

**Figure S9 | Images and characterization of the MoS₂ flake used for the micro-lens before FIB milling. a,** Optical microscope image of the MoS₂ flake used for the micro-lens, before FIB milling. **b,** PSI image of the MoS₂ flake used for the micro-lens, before FIB milling, from the box within the dashed line indicated in **(a)**. **c,** OPL measured by PSI along the dash line in **(b)**. **d,** AFM image from the box within the dashed line indicated in **(b)**.

The 9L MoS₂ flake (Figure S9) was fabricated into a micro-lens with the FEI FIB system. A Gallium ion source (30 kV voltage, 9.7 pA current) was used in the FIB milling process. The micro-lens parameters, such as the diameter and the depth, are from our simulation pre-design.

## 6.2 Focal length calculation based on the measured OPL profile in experiment

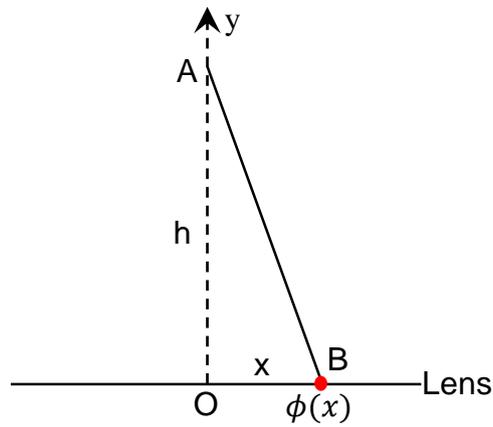

**Figure S10 | Schematic plot of the focal length calculation for the MoS$_2$ micro-lens.**

Figure S10 shows the schematic cross section of the lens. The phase shift $\phi(x)$ is impinged on the $x$ axis, where point $O$ is the center of the lens. The focal length of the micro-lens can be extracted based on the phase shift $\phi(x)$ profile measured in experiment. We evaluate the interference of light along the perpendicular axis $y$. For each point on the lens plane, we calculated the phase shifts, including the propogation phase and the phase induced by the MoS$_2$ lens. For example, for a point B, the point A receives light from a point B with an phase as $e^{i\phi(x)}e^{-i2\pi(h^2+x^2)/\lambda}$. The minus sign in the exponential term is due to the fact that the phase shift profile produces a virtual focus. We can find the focal length by evaluating the maximum value of the following term:

$$I(y) = \int_0^r dx\, e^{i\phi(x)}e^{-i2\pi(h^2+x^2)/\lambda} \tag{S27}$$

where $r$ is the radius of the lens. Using the experimentally measured OPL profile as shown in Figure 3d, we calculated the focal length of the MoS$_2$ micro-lens to be -248 μm.

## 6.3 Micro-lens characterization details

We exploited the light path of the Horiba Jobin Yvon T64000 micro-Raman system and set up a far-field scanning optical microscope (SOM). The schematic plot of far-field SOM is shown in Figure S11. The SOM system used a green laser (at 532 nm) that was focused by an Olympus 10X (NA = 0.25, depth of focus 18 μm) objective lens. If the sample was placed in the focal plane of the objective, the light coming from the focal spot on the sample converged into a light spot at the camera. In these conditions the light path is indicated by the solid line in Figure S11. Otherwise, if the sample is out of the focal plane, the light coming the sample will be either over-focused into a spot in front of the camera or less-focused, and leave a halo at the camera, as indicated by two sets of dashed lines in Figure S11. The setup offers good collection efficiency for the light coming from (or effectively coming from) a small volume around the focal plane.

In the characterization experiment, the MoS$_2$ micro-lens was moved along the $z$ axis direction in steps of 10 μm by a piezo-electrically driven stage. The camera recorded a series of the intensity distributions (Figure S12), when the MoS$_2$ micro-lens was put at different $z$ values. For comparison, we also ran the same characterization by using a planar substrate without the MoS$_2$ micro-lens, and obtained a similar set of intensity distribution images (Figure S13). The lensing effect is clearly demonstrated by comparing the difference between Figure S12 and S13.

The characterization principle for the MoS$_2$ micro-lens is shown in Figure S12. Our MoS$_2$ micro-lens is effectively a (reflective) concave lens and we can use following lens formula to characterize the relationship among the focal length $f$, objective distance $f_1$ and image distance $f_2$.

$$\frac{1}{f} = \frac{1}{f_1} + \frac{1}{f_2} \tag{S28}$$

For this MoS$_2$ micro-lens, the focal length $f$ has a negative value. In our experimental characterization, $f_1$ has a negative value and its absolute value is always the distance between the focal plane and the lens, $d$. With $f_1 = -d$, equation S28 can be rewritten as:

$$\frac{1}{f_2} = \frac{1}{d} - \frac{1}{|f|} \tag{S29}$$

Based on equation S29, we can determine $f_2$ and the corresponding light paths in five cases.

When $0 < d < |f|$, from equation S29 we can get that $f_2 > 0$, which means that a real light spot will be formed on the upper side of the MoS$_2$ micro-lens, as indicated in Figure S12a left panel. Under this condition, the camera will observe a circular disk pattern that is effectively coming from the focal plane. The observed intensity pattern distribution by camera at $z = -120$ µm, is shown in Figure S12a right panel.

When $d = |f|$, from equation S29 we can get that $f_2 = \infty$, which means that the light will be reflected back as a group of parallel light, as indicated in Figure S12b left panel. The observed intensity pattern distribution by camera at $z = -240$ µm, is shown in Figure S12b right panel.

When $|f| < d < 2|f|$, from equation S29 we can get that $f_2 < -2|f|$, which means that lights will be reflected as indicated in Figure S12c. The observed intensity pattern distribution by camera at $z = -360$ µm, is shown in Figure S12c right panel.

When $d = 2|f|$, from equation S29 we can get that $f_2 = -d = -2|f|$, which means that lights will be reflected exactly back along the same route of the incident light by the micro-lens, as indicated in Figure S12d. The observed intensity pattern distribution by camera at $z = -480$ µm, is shown in Figure S12d right panel.

When $d > 2|f|$, from equation S29 we can get that $-2|f| < f_2 < -|f|$, which means that lights will be reflected back as indicated in Figure S12e. The observed intensity pattern distribution by camera at $z = -600$ µm, is shown in Figure S12e right panel.

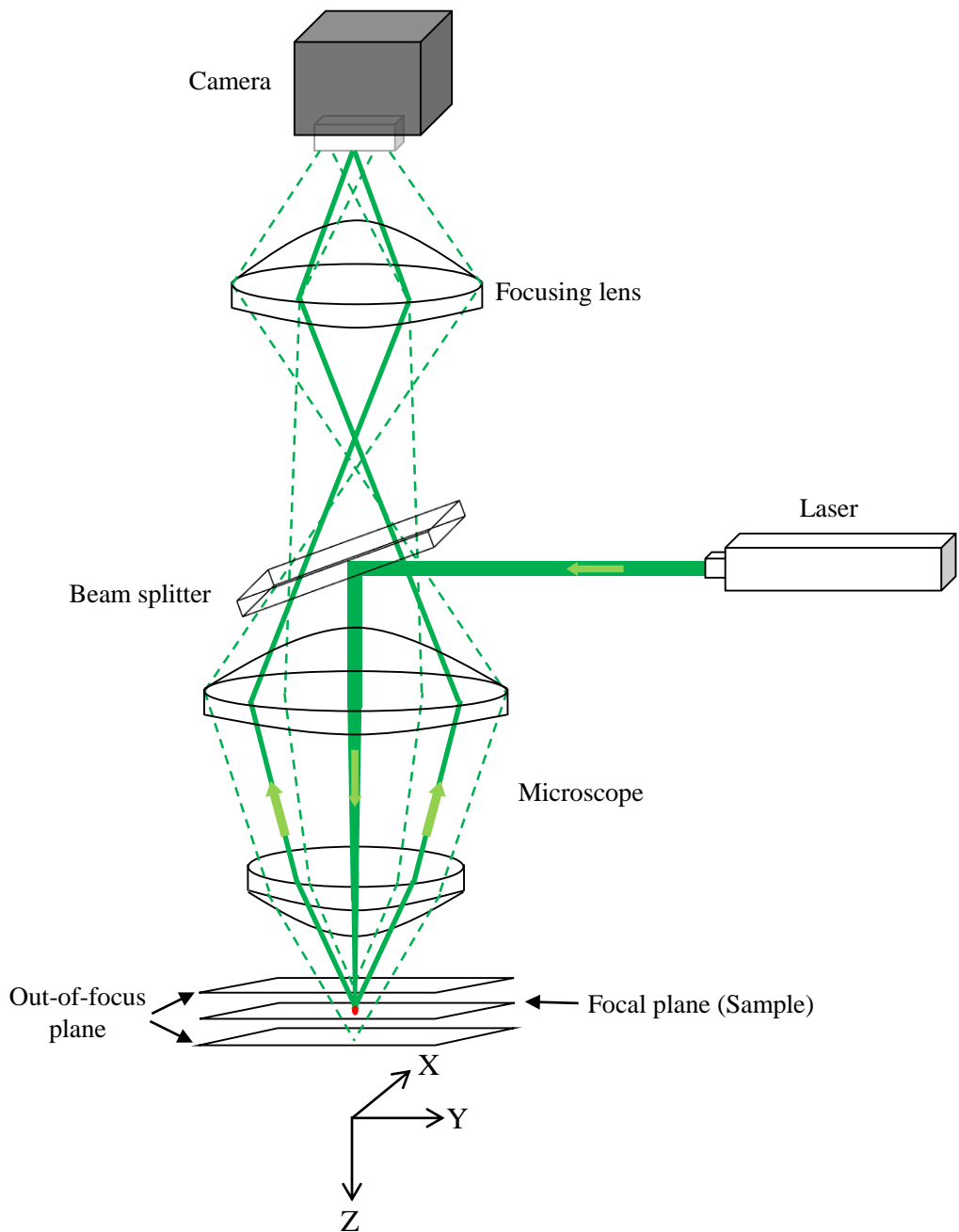

**Figure S11 | Schematic plot of the far-field scanning optical microscope (SOM) used for MoS$_2$ micro-lens characterization.**

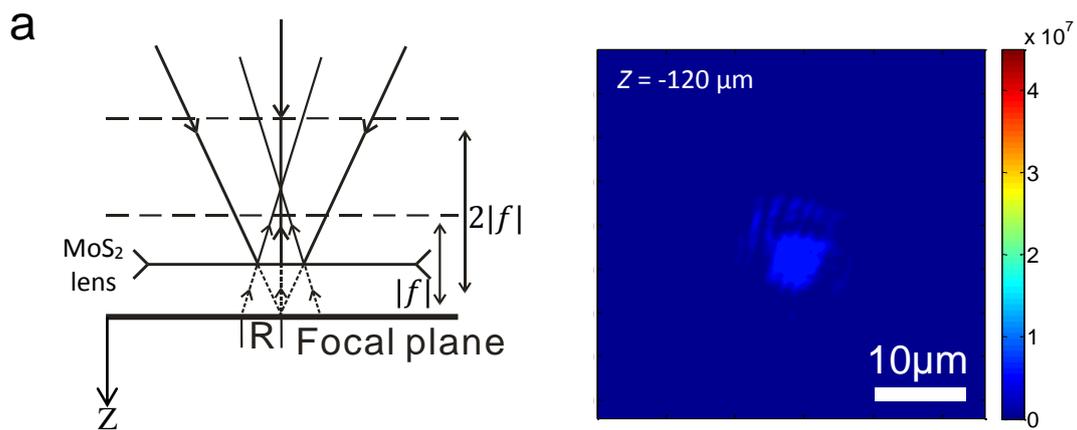

a

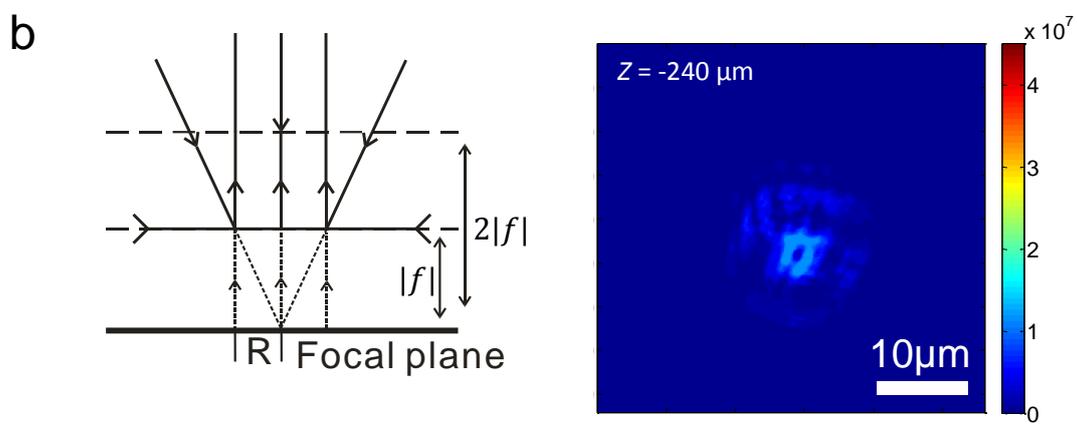

b

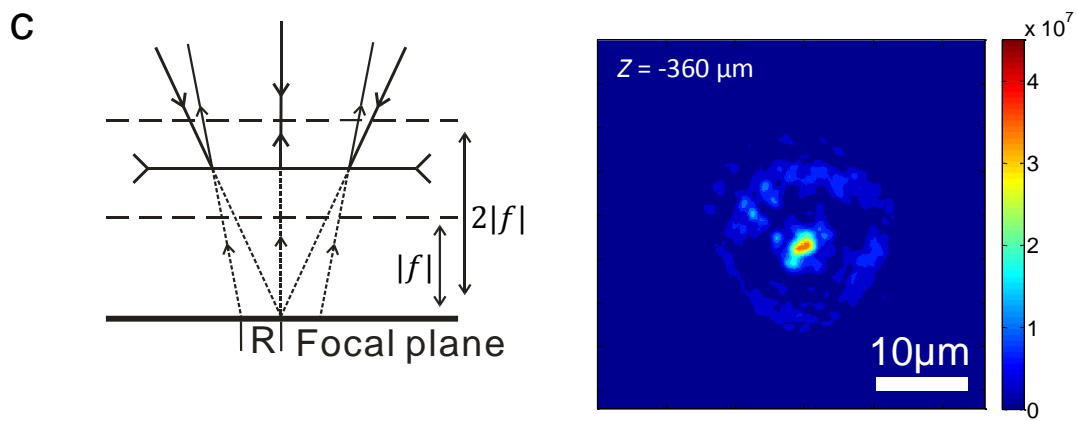

c

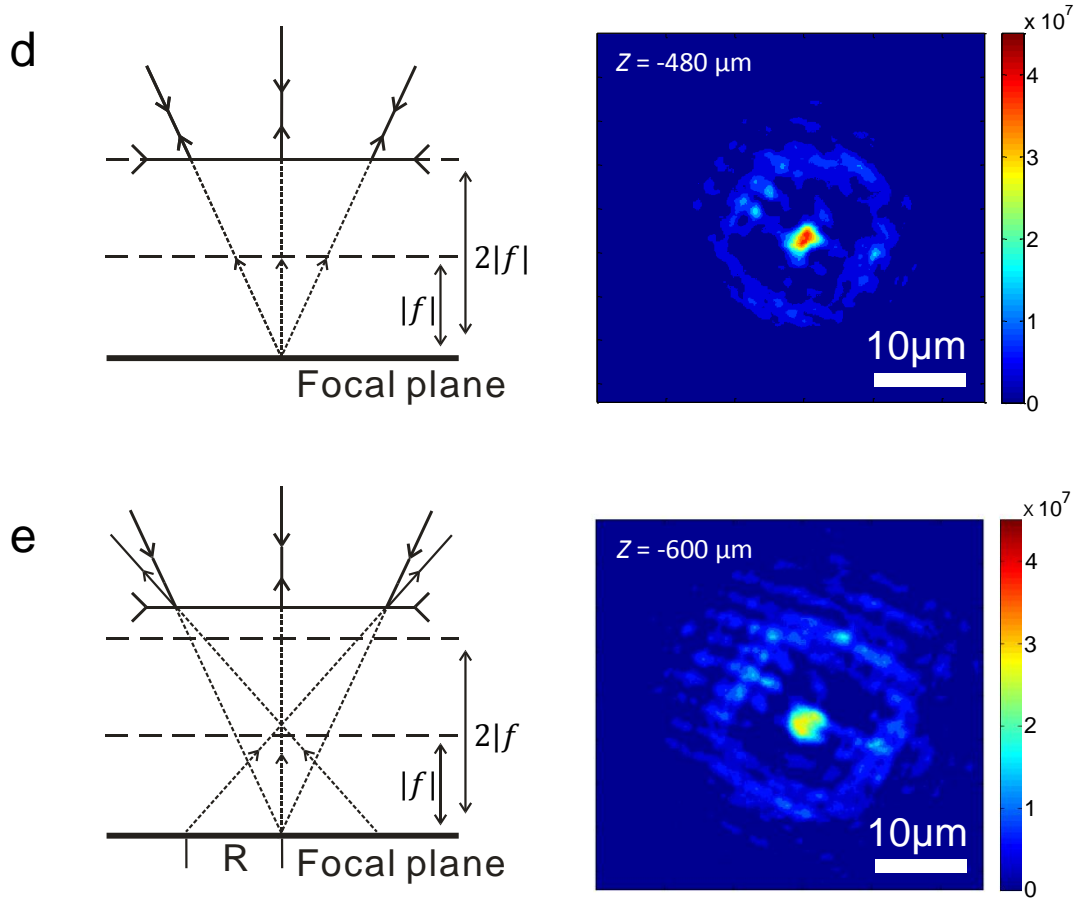

**Figure S12 | Characterization of the atomically thin MoS₂ micro-lens. a-e,** Schematic plots of the light path (left) and the corresponding recorded intensity distribution pattern images (right) by camera, when **(a)** $d < |f|$, **(b)** $d = |f|$, **(c)** $|f| < d < 2|f|$, **(d)** $d = 2|f|$ and **(e)** $d > 2|f|$. Note: *d* is the distance between the micro-lens and the focal plane, a positive value; *f* is the focal length of the micro-lens.

For comparison, we ran a control measurement on the planar SiO₂/Si substrate using the same procedure. The schematic plots and recorded images are indicated in Figure S13. The SiO₂/Si substrate was considered to be a flat mirror. In Figure S13, three conditions are illustrated: $d < |f|$, $|f| < d < 2|f|$ and $d > 2|f|$. In the control measurements, when the distance between the planar substrate and the focal plane $d$ increased, the radius of the circular disk pattern observed by the camera will increase, without any lensing effect. The fringe patterns is the

SOM system background noise, which is due to the interference from light beams reflected by various interfaces in the optics.

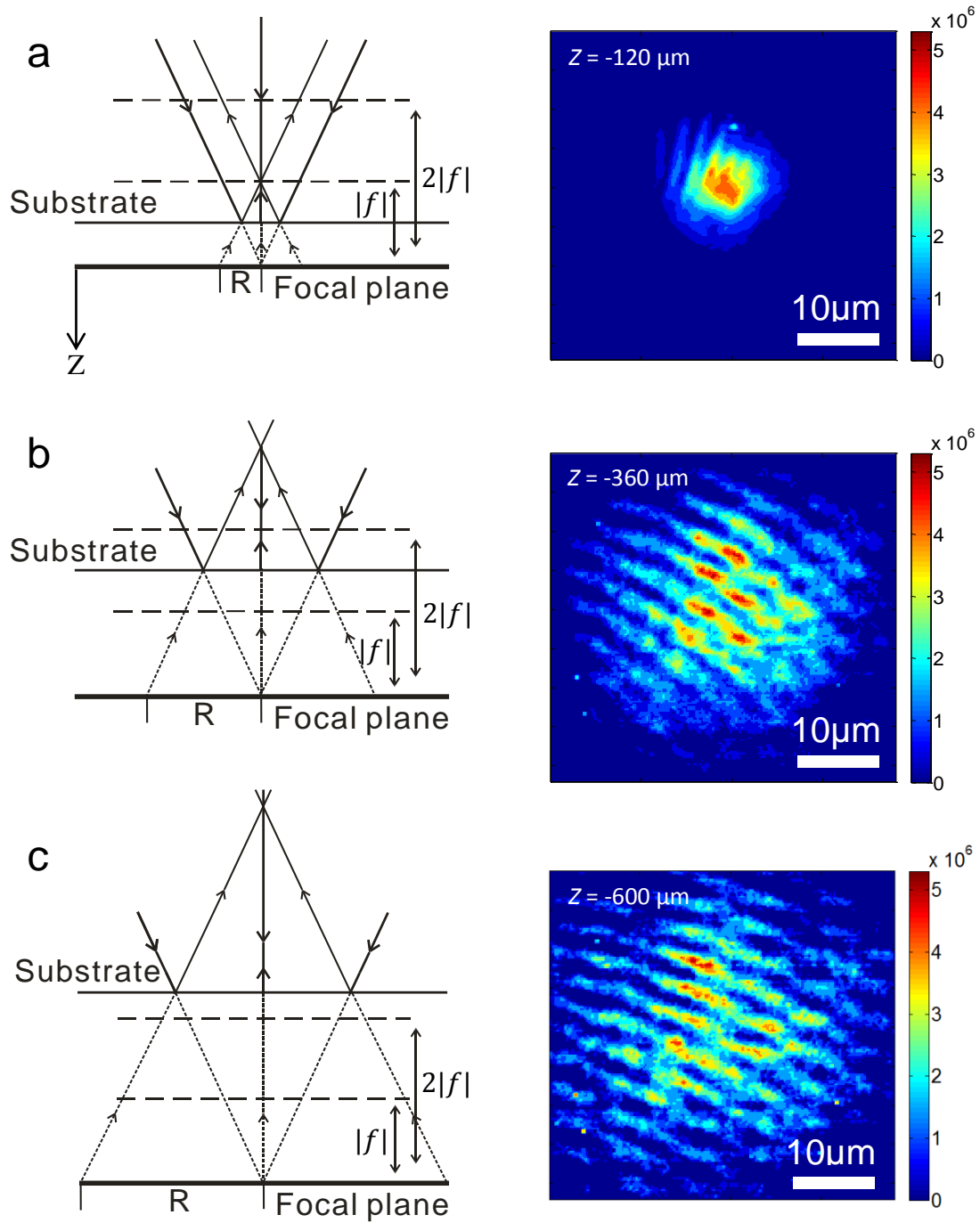

**Figure S13 | Characterization control experiments using a planar SiO$_2$ substrate. a-c,** Schematic plots of the light path (left) and the corresponding camera recorded optical images (right), when **(a)** $d < |f|$, **(b)** $|f| < d < 2|f|$ and **(c)** $d > 2|f|$. Note: $d$ is the distance between the micro-lens and the focal plane; $f$ is the focal length of the MoS$_2$ micro-lens.

In the experiment, we kept the same laser power for the whole measurement. But as stage moves upwards (*z* absolute value increases), the laser spot size on the stage (micro-lens plane) will increase and the power density will decrease appropriately following the scaling of $1/z^2$. In order to make sure that the incident beam has the same power area density at all *z* values, we made the following normalization for all the recorded images:

$$Intensity_{normalized} = Intensity_{measured} \times z^2 \qquad (S30)$$

where $Intensity_{normalized}$ is the normalized intensity, $Intensity_{measured}$ is the intensity of the spot in the image, and *z* is the value of height where the image was taken. Then a three dimensional dataset was composed of the series of the images along *z* axis. A cross section profile was obtained along the *x*- and *z*-axes to illustrate the distribution of the intensity along these directions. To better represent the data acquired during the experiment, the intensities of the cross section along *x*- and *z*-axes were normalized. At a given distance, a virtual circle was drawn to the center peak of the spot, and 600 points were chosen evenly on this circle and their intensities were averaged. Then the average intensity was selected to represent the value at this specific radius to the center. Finally, the data at different heights were assembled, interpolated and plotted in contour. All the images in Figure 3e & 3f and images in Figure S12 & S13 are based on the normalized intensities.

### 7. Polarization-dependence characterization on $MoS_2$ micro-lens

A linearly polarized laser (532 nm wavelength, at normal incidence) was used to characterize the focal length of our $MoS_2$ micro-lens. It shows that the measured focal length has very weak polarization dependence (Figure S14a). This is because $MoS_2$ has weak anisotropic dielectric response. This is consistent with that our measured photoluminescence (PL) from monolayer $MoS_2$ has very weak polarization dependence (Figure S14b).

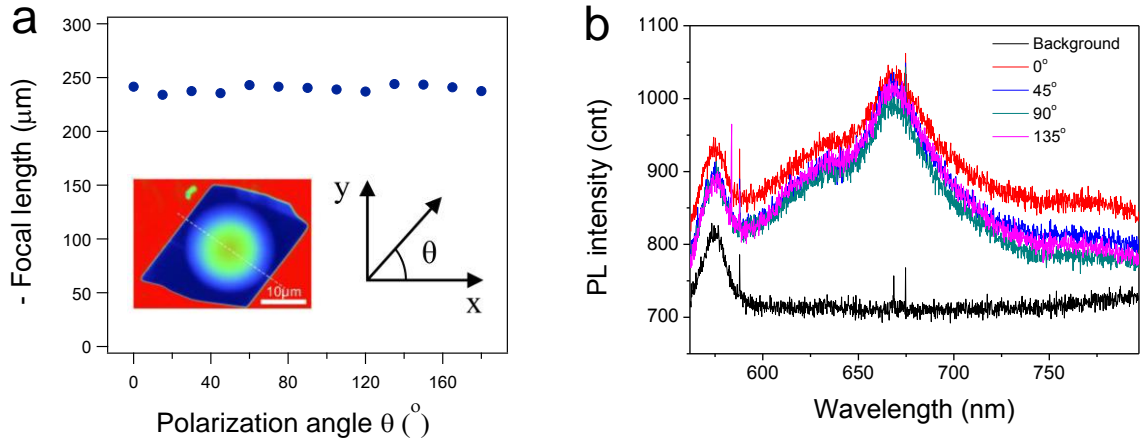

**Figure S14 | a**, Measured focal length polarization dependence of our MoS$_2$ micro-lens. **b**, Measured PL polarization dependence on a monolayer MoS$_2$ sample. Linearly polarized 532 nm laser at normal incidence was used as the excitation.

## 8. Calculation and experiment details for atomically thin grating

### 8.1 Simulations for scattering cross section

The scattering cross section of a 30 nm wide ribbon was calculated using a finite element method, which solves the full-wave Maxwell's equations numerically. The simulation was performed in two-dimensional space. The cross section was calculated for different incident angles, showing excellent isotropic response for s-polarized light (Figure S15).

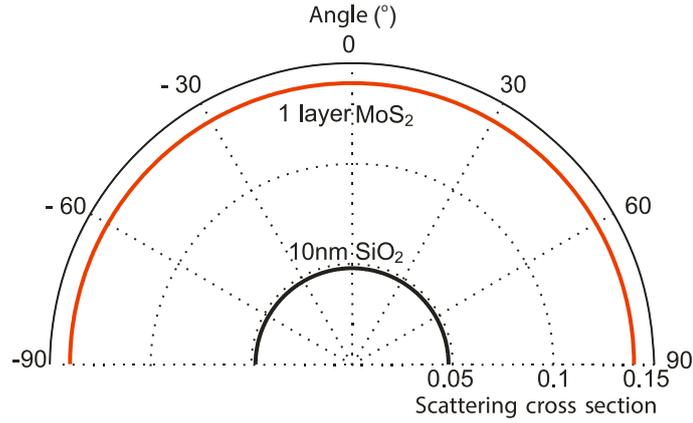

**Figure S15 | Isotropic scattering cross section of different materials.** The scattering cross section for a 1L MoS$_2$ (red line) nano-ribbon with a 30 nm width for light illumination from different incident angles. The angular response is isotropic. In comparison, the black line indicates the scattering cross section for a 10 nm thick SiO$_2$ ribbon with a 30 nm width.

**8.2 Grating fabrication and characterization**

Gratings on thin-layers (1L, 2L, 5L, 6L and 8L) of MoS$_2$ (Figure S16, S17 and S18), control gratings on a SiO$_2$ substrate, and on a monolayer of graphene (Figures S19) were all fabricated using an FEI focussed ion beam (FIB) system. A Gallium ion source was used in the milling process. The grating parameters, such as periodicity, 2D material's filling ratio, were based on our simulation results. A grating diffraction efficiency measurement setup was established, as indicated in Figure 4b. The sample chip with the grating on top was mounted onto a turnplate. The grating could be rotated and make sure the incident laser beam was always in the plane that is normal to the line grating. The incident parallel laser beam (532 nm wavelength) is s-polarized and has a diameter of around 200 μm, which fully covers the grating. A power meter was used to measure the power of the incident laser and diffraction light. In the output power measurements, the light was always perpendicular to the power meter. To calculate the grating diffraction efficiency, the following formula was used:

$$\eta = \frac{P_d}{P_i} * \frac{S_b}{S_g} \tag{S31}$$

where $\eta$ is the grating diffraction efficiency, $P_d$ is the diffraction light power, $P_i$ is the incident laser power, $S_b$ is the area size of the laser beam and $S_g$ is the area size of the grating.

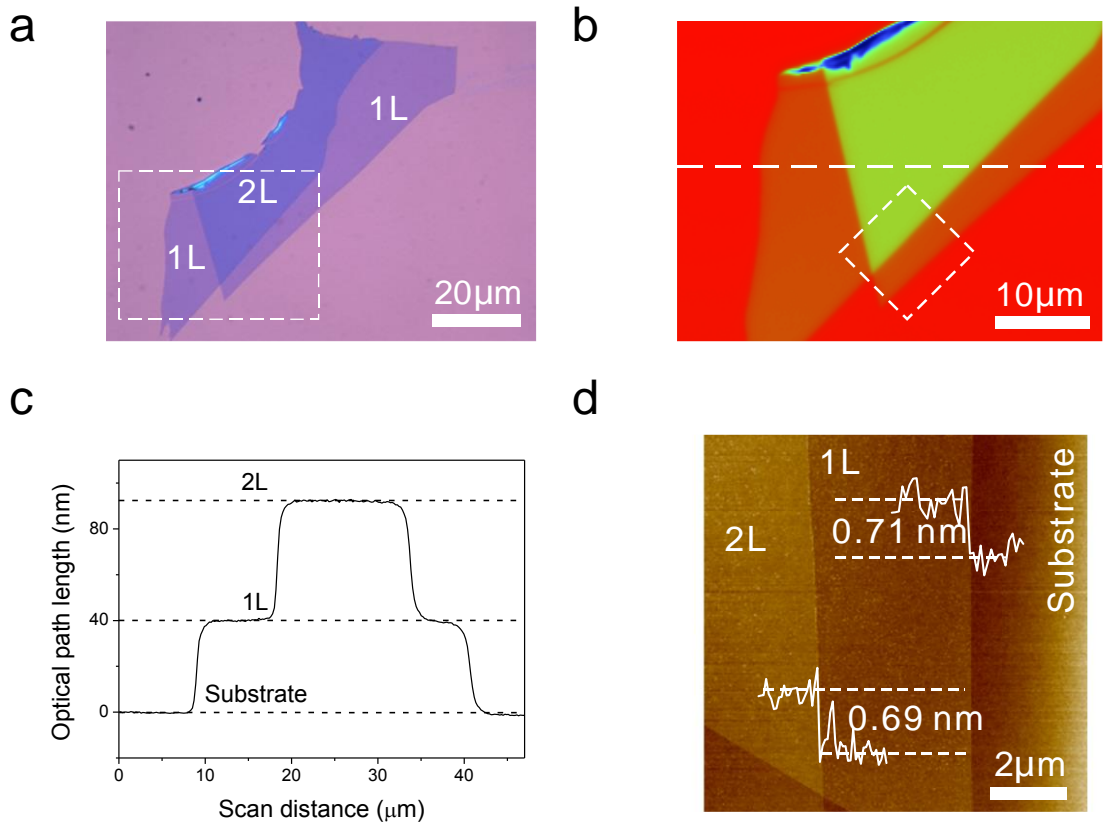

**Figure S16 | Images and characterization of 1L and 2L MoS$_2$ flake before grating FIB milling. a,** Optical microscope image of 1L and 2L MoS$_2$ before FIB milling. **b,** PSI image of 1L and 2L MoS$_2$ before FIB milling, from the dash line box area indicated in **(a)**. **c,** OPL measured by PSI along the dash line indicated in **(b)**. **d,** AFM image from the dash line box indicated in **(b)**.

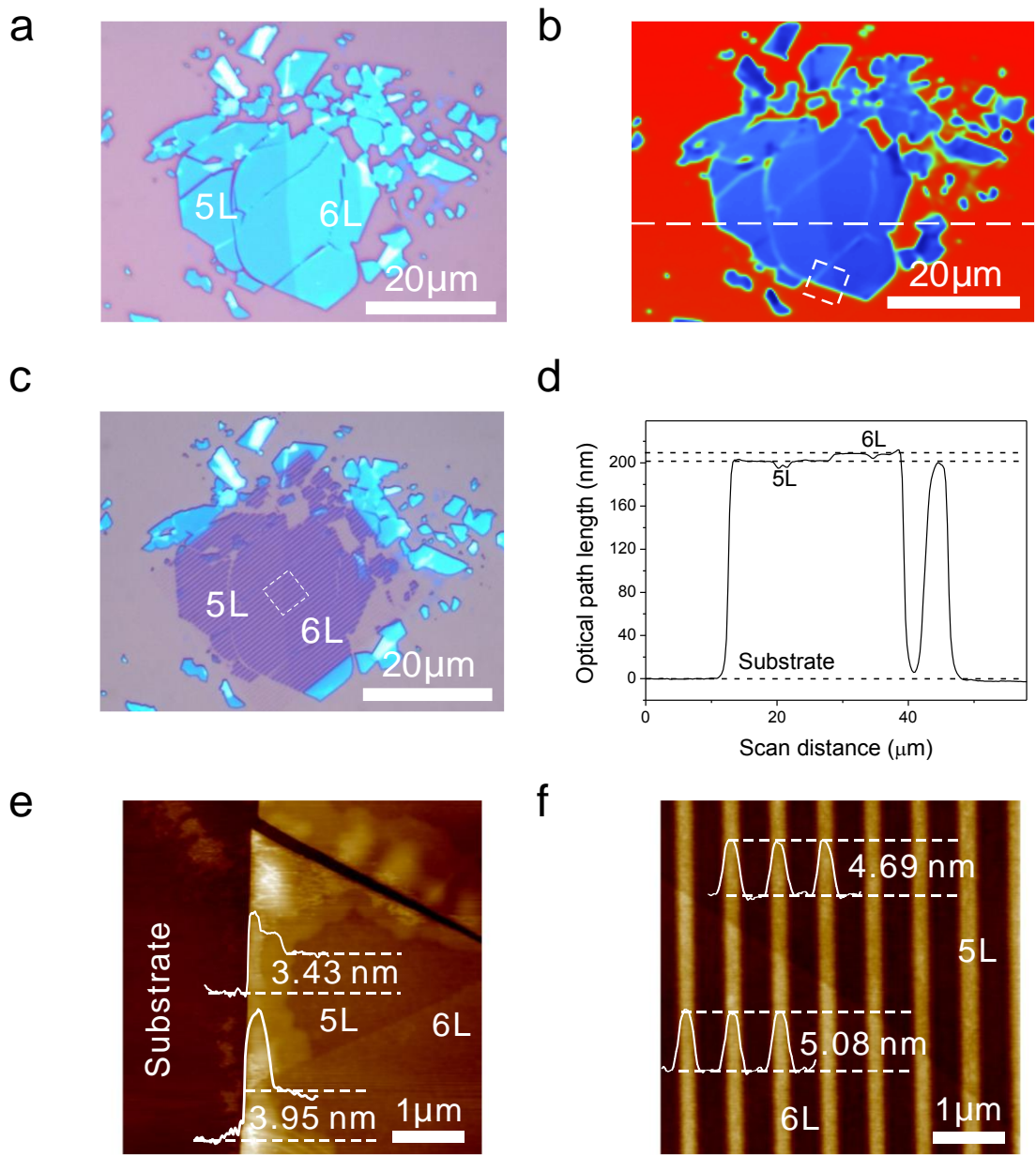

**Figure S17 | Images and characterization of 5L and 6L MoS₂ grating. a,** Optical microscope image of 5L and 6L MoS$_2$ flake before FIB milling. **b,** PSI image of 5L and 6L MoS$_2$ before milling. **c,** Optical microscope image of 5L and 6L grating. **d,** OPL measured by PSI along the dash line indicated in (**b**). **e,** AFM image from the dash line box indicated in (**b**). **f,** AFM image of 5L and 6L grating from the dash line box in (**c**). Note: based on the measured grating height, the 5L and 6L MoS$_2$ were fully etched through and the SiO$_2$ substrates underneath were over etched by around 1.2 nm. From our control SiO$_2$ grating experiment and simulation, the grating contribution from this over etched SiO$_2$ is negligible.

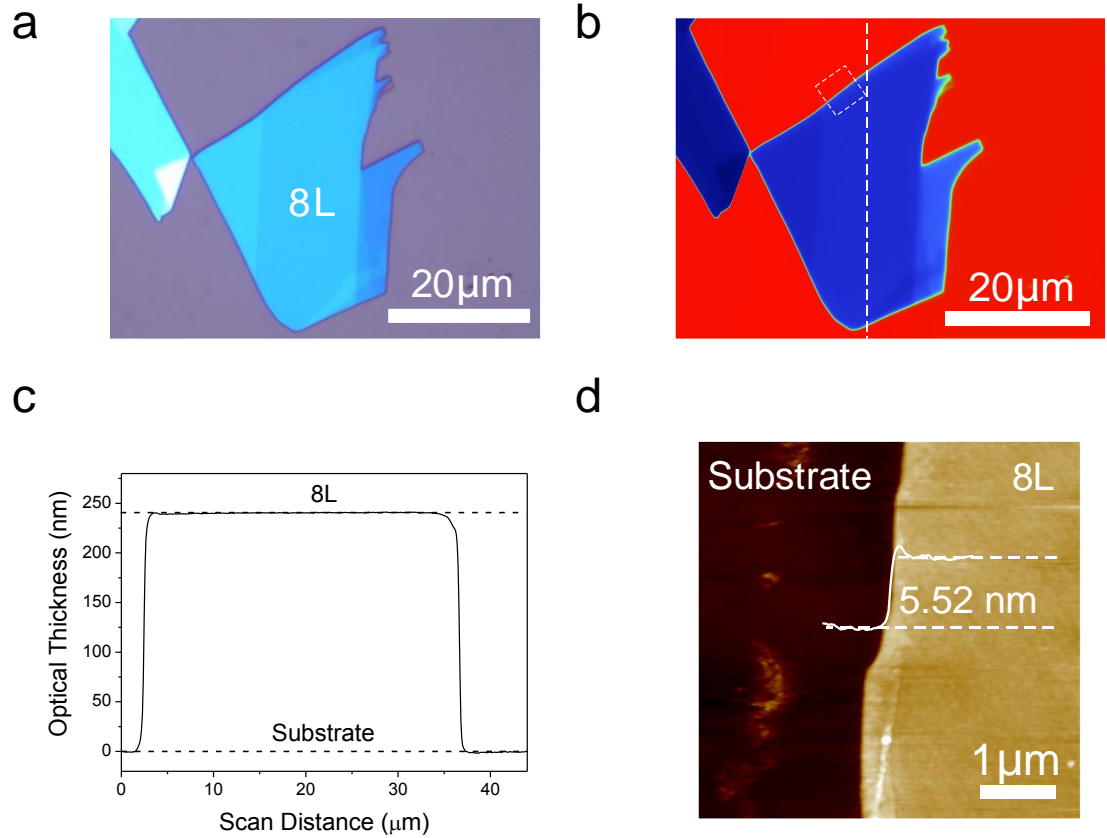

**Figure S18 | Images and characterization of 8L MoS$_2$ before grating FIB milling. a,** Optical microscope image of 8L MoS$_2$ before milling. **b,** PSI image of 8L MoS$_2$ before milling. **c,** OPL measured by PSI along the dash line indicated in **(b)**. **d,** AFM image from the dash line box indicated in **(b)**.

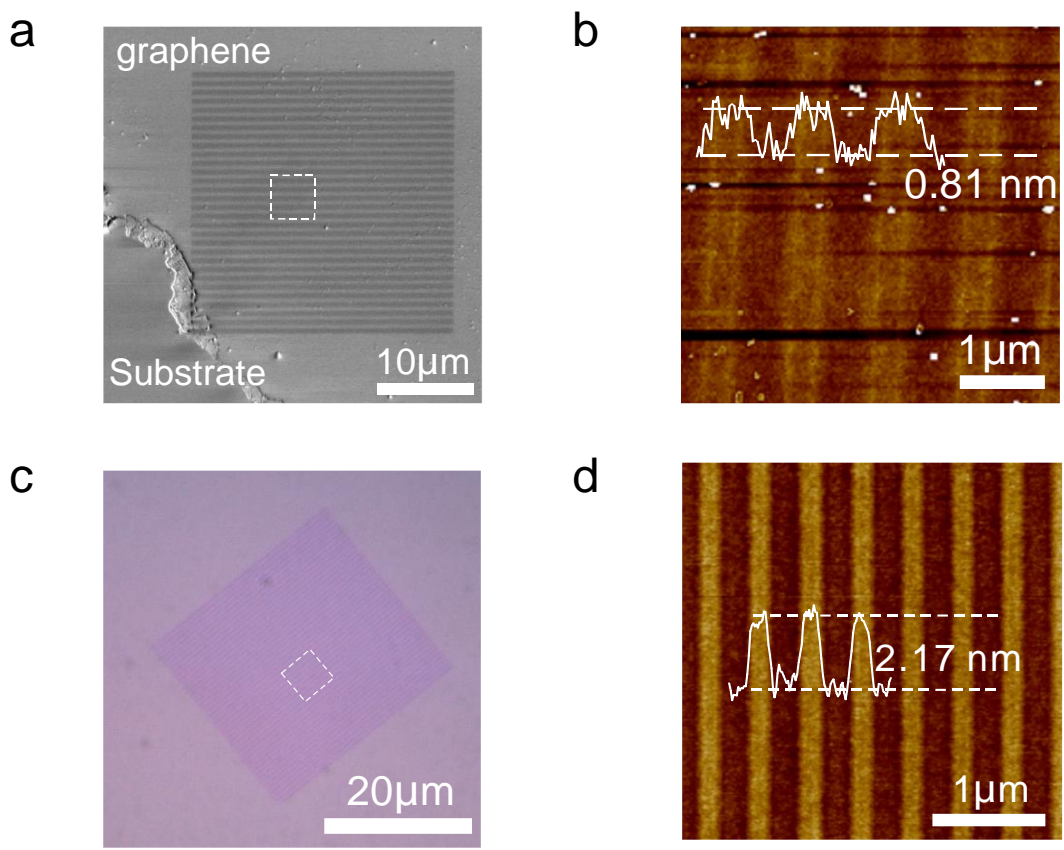

**Figure S19 | Images and characterization of control gratings. a,** SEM image of the control grating on graphene. The single-layer graphene used in the control grating was deposited by chemical vapour deposition (CVD) on copper substrate at 1000 ºC and then transferred to Si/SiO$_2$ substrate, as reported[11]. **b,** AFM image of the control grating on graphene. Note: based on the measured height, the graphene was fully etched through and the SiO$_2$ substrate underneath was over etched by around 0.45 nm. **c,** Optical microscope image of the control grating on SiO$_2$. **d,** AFM image of the control grating on SiO$_2$.

## 8.3 Calculation details for grating efficiencies

The grating efficiencies were calculated using rigorous coupled wave analysis (RCWA). Three parameters were scanned in order to search the maximum efficiency: the periodicity, the filling ratio of MoS$_2$, and the incident angle. The optimal efficiencies and corresponding parameters are given in Table S1. In our simulations, we used the reported refractive index values for

$MoS_2$[10] and graphene[12], which are n = 5.3 + 1.3i and n = 2.6 + 1.3i, respectively. The optimal efficiency and the parameters are listed in table S2. Figure S21 shows the 1st order diffraction efficiency for 8 layer $MoS_2$. The efficiency was calculated as a function of periodicity and incident angle for a fixed filling ratio.

The grating efficiency can be further improved with a reflective mirror behind the $MoS_2$ grating. Here we calculated the efficiency of the 1st order diffraction beam from an 8-layer $MoS_2$ grating. The layer structure is shown in Figure S20. The thickness of $SiO_2$ is 223 nm, which corresponds to the second Fabry-Perot resonance for the $SiO_2$ layer. We performed a three-dimensional parameter scan including the filling ratio of $MoS_2$, the periodicity, and the incident angle.

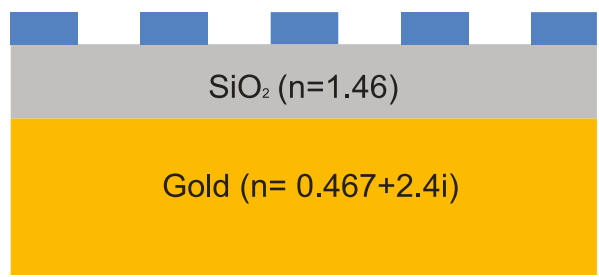

**Figure S20 | Layer Structure of the $MoS_2$ grating with an Au reflective mirror.**

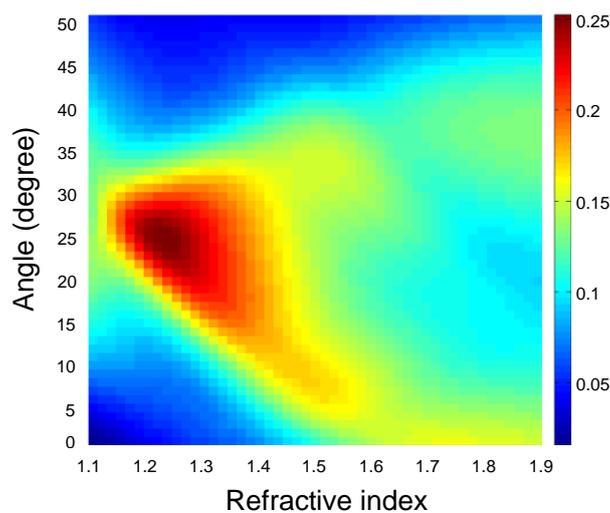

**Figure S21 | 1st order diffraction efficiency map for an 8-layer $MoS_2$ grating with an Au reflective mirror.**

**Table S1 Optimal parameters and 1st order efficiency for gratings with $SiO_2$/Si substrate.**

| Materials | Filling ratio | Periodicity | Incident angle (º) | Efficiency-simulation (%) | Efficiency-experiments (%) |
|---|---|---|---|---|---|
| 1L $MoS_2$ | 0.52 | 1.8λ | 8.7 | 0.4 | 0.3 |
| 2L $MoS_2$ | 0.39 | 1.2λ | 23.9 | 1.3 | 0.8 |
| 6L $MoS_2$ | 0.39 | 1.2λ | 23.9 | 7.4 | 4.4 |
| 8L $MoS_2$ | 0.44 | 1.2λ | 23.9 | 10.2 | 10.1 |
| 2 nm $SiO_2$ | 0.44 | 1.2λ | 21.0 | 0.0051 | - |
| 1L graphene | 0.52 | 1.9λ | 8.0 | 0.0078 | - |
| 1 nm Au | 0.49 | 1.9λ | 7.8 | 0.0519 | - |

**Table S2 Optimal parameters and 1st order efficiency for 8L $MoS_2$ grating with an Au reflective mirror.**

| Material | Filling ratio | Periodicity | Incident angle (º) | Efficiency (%) |
|---|---|---|---|---|
| 8L $MoS_2$ | 0.43 | 1.5λ | 23.5 | 23.7 |

## 9. Measured giant OPL in WS$_2$ and WSe$_2$

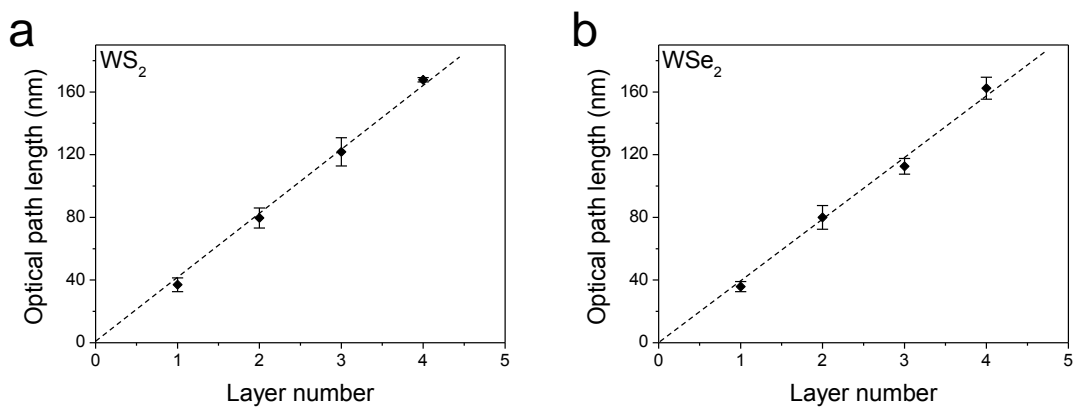

**Figure S22 | Statistical OPL data from single- to quadri-layer WS$_2$ (a) and WSe$_2$ (b), measured by PSI.** At least five samples were measured for each layer number and all the layer numbers were confirmed by AFM.